\documentclass[12pt]{article}
\usepackage{epsfig,amssymb,cite,multirow} 
\usepackage{amsmath}
\def\calg         {{\cal G}}

\def\caln         {{\cal N}}

\def\calw         {{\cal W}}

\def\tr           {\mathop{\rm Tr}}

\def\Im           {{\rm Im\hskip0.1em}}

\def\d         {{\rm d}}

\def\sqr#1#2{{\vcenter{\vbox{\hrule height.#2pt
 \hbox{\vrule width.#2pt height#1pt \kern#1pt \vrule width.#2pt}\hrule
 height.#2pt}}}}

\makeatletter
\@addtoreset{equation}{section}
\makeatother

\def\beq{\begin{equation}}                     %
\def\eeq{\end{equation}}                       %
\def\bea{\begin{eqnarray}}                     
\def\eea{\end{eqnarray}}   
\def\nn{\nonumber}                    
              
\begin{document}

\setcounter{page}{0}
\begin{titlepage}
\titlepage
{\scriptsize\rightline{CERN-PH-TH/2007-235}
\rightline{BICOCCA-FT-07-16 }
\rightline{SISSA 87/2007/EP}
\rightline{LMU-ASC 69/07}
}

\vskip 1.5cm
\centerline{{ \bf \Large On the geometry and the moduli space}} 
\vskip 0.5cm
\centerline{{ \bf \Large of  $\beta$-deformed quiver gauge theories }}
\vskip 1.5cm
\centerline{Agostino Butti$^{a}$, Davide Forcella$^{b}$, Luca Martucci$^{c}$,} 
\centerline{Ruben Minasian$^{d}$, Michela Petrini$^{a}$
and Alberto Zaffaroni$^e$}
\begin{center}
$^a$ LPTHE, Universit\'es Paris VI, Jussieu \\
75252 Paris, France
\vskip .2cm
$^b$ International School for Advanced Studies (SISSA/ISAS) \\
and \\
INFN-Sezione di Trieste,\\
via Beirut 2, I-34014, Trieste, Italy\\
\vskip .2cm
$^b$ PH-TH Division, CERN CH-1211 Geneva 23, Switzerland\\
\vskip .2cm
$^c$ Arnold Sommerfeld Center for Theoretical Physics, LMU M\"unchen,\\
Theresienstra\ss e 37, D-80333 M\"unchen, Germany
\vskip .2cm
$^d$Service de Physique Th\'eorique,                   
CEA/Saclay \\
91191 Gif-sur-Yvette Cedex, France  
\vskip .2cm

$^e$ Dipartimento di Fisica, Universit\`a di Milano Bicocca and INFN \\
sezione Milano-Bicocca, piazza della Scienza 3, Milano 20126, Italy
\end{center}
\vskip 1.5cm  
\begin{abstract}
  We consider a class of super-conformal $\beta$-deformed $\mathcal
  N=1$ gauge theories dual to string theory on $AdS_5 \times X$ with
  fluxes, where $X$ is a deformed Sasaki-Einstein manifold. 
The supergravity backgrounds are explicit examples of Generalised Calabi-Yau manifolds:
the cone
  over $X$ admits an integrable generalised complex structure in terms
  of which the BPS sector of the gauge theory can be described. The
  moduli spaces of the deformed toric $\mathcal N=1$ gauge theories
  are studied on a number of examples and are in agreement with the
  moduli spaces of D3 and D5 static and dual giant probes.
\end{abstract}

\end{titlepage}

\newpage

\section{Introduction}

The super-conformal gauge theories living on D3-branes at
singularities generally admit marginal deformations. A particularly
interesting case of marginal deformation for theories with $U(1)^3$
global symmetries is the so called $\beta$-deformation \cite{LS}.  The
most famous example is the $\beta$-deformation of ${\cal N}=4$ SYM which
has been extensively studied both from the field theory point of view
and the dual gravity perspective.  In
particular, in \cite{LM}, Lunin and Maldacena found the supergravity dual solution, which is
a  completely regular $AdS_5$ background.
Their construction can be generalised to the 
super-conformal theories
associated with the recently discovered Sasaki-Einstein backgrounds
$AdS_5\times L^{p,q,r}$ \cite{YL}.  More generally, 
all toric quiver gauge theories admit $\beta$-deformations \cite{HB}
and, as we will see, have regular gravitational duals. The
resulting $\beta$-deformed theories are interesting both from the
point of view of field theory and of the gravity dual.

On the
field theory side, we deal with a gauge theory with a deformed moduli
space of vacua and a deformed spectrum of BPS operators. The case of
${\cal N}=4$ SYM has been studied in details in the literature
\cite{bl,dorey,benini}. In this paper we extend this analysis to
a generic toric quiver gauge theory. The moduli space of the $\beta$-deformed
gauge theory presents the same features as in ${\cal N}=4$ case. In particular,
its  structure depends on the value of the deformation parameter $\beta$. 
For generic $\beta$ the deformed theory admits a Coulomb branch which is given by  
a set of complex lines. For $\beta$ rational there are additional directions corresponding
to Higgs branches of the theory.

On the gravity side, the dual backgrounds can be obtained from the original Calabi-Yaus
with a continuous T-duality transformation using the general method proposed in \cite{LM}.
We show that it is possible to study the $\beta$-deformed background
even in the cases where the explicit original Calabi-Yau metric is not known. The toric structure 
of the original  background is enough. Besides the relevance for AdS/CFT, 
the $\beta$-deformed backgrounds are also interesting from the geometrical point view.
They are Generalised Calabi-Yau manifolds\cite{gmpt1,gmpt2}:  after the
deformation the background is no longer complex, but it still admits an integrable 
generalised complex structure.
Actually the $\beta$-deformed backgrounds represent  one of the few explicit known
examples of generalised geometry solving the equation of motions of
type II supergravity \footnote{For other non compact examples see \cite{bgmpz,mpz} 
and for compact ones \cite{GMPT3}.}.  
The extreme simplicity of such backgrounds make it possible to explicitly apply the formalism
of Generalised Complex Geometry, which, as we will see, provides an elegant way to
study T-duality and brane probes \cite{paul,Martucci,lucasup,lucapaul1}.

The connection between gravity and field theory is provided by the
study of supersymmetric D-brane probes moving on the $\beta$-deformed
background. 
In this paper we will analyse the case of static D3 and D5
probes, as well as the case of D3 and D5 dual giant gravitons. 
We will study in details existence and moduli space of such probes.
We show that, in the $\beta$-deformed background, both 
static D3 probes and D3 dual giants can only live on a set of intersecting 
complex lines inside the
deformed Calabi-Yau, corresponding to the locus where the $T^3$ toric
fibration degenerates to $T^1$.  This is in agreement with the abelian
moduli space of the $\beta$-deformed gauge theory which indeed consists
of a set of lines.  Moreover, in the case of rational $\beta$, we
demonstrate the existence of both static D5 probes and D5 dual giant gravitons
with a moduli space isomorphic to the original Calabi-Yau divided by a
$\mathbb{Z}_n\times \mathbb{Z}_n$ discrete symmetry.  This statement
is the gravity counterpart of the fact that, for rational $\beta$, new
branches are opening up in the moduli space of the gauge theory
\cite{bl,dorey}.
Our analysis also generalises the results of \cite{MSg} where it has been shown
that the classical phase space
 of supersymmetric D3 dual giant gravitons in the undeformed
Calabi-Yau background is isomorphic to the Calabi-Yau variety.

The classical way to study probe configuration is to solve the equations of motion
coming from the probe Dirac-Born-Infeld action.
Generalised Complex Geometry provides an alternative 
method to approach the problem. As we will explain, a D-brane is characterised by its generalised
tangent bundle. The dual probes in the
$\beta$-deformed geometry can be obtained from the original ones applying T-duality 
to their generalised tangent bundles. 
The approach in terms of Generalised Geometry allows also to clarify how the complex 
structure of the gauge theory is
reflected by the gravity dual, which, as we have already mentioned,
is not in general a complex manifold.

The study of brane probes we present here can be seen as consisting of two
independent and complementary sections, 
one dealing with the Born-Infeld approach and
the other one using  Generalised Complex Geometry.  
We decided to keep the two analysis independent, so that the reader not
interested in one of the two can skip the corresponding section.

The paper is organized as follows. In Section 2 we discuss the
structure of the $\beta$-deformed gauge theory and of its gravity dual,
and we characterize it in terms of pure spinors.  
In Section 3 we study the moduli space of D3 and D5-brane, static probes and dual giant gravitons,
on the deformed background using the Born-Infeld action, while
in Section 4 we analyse the same configurations using the generalised tangent bundle approach.
We will show that, as usual
for BPS quantities, the explicit knowledge of the Calabi-Yau metric is
not required to extract sensible results.  Our analysis thus applies
to the most general toric background. In Section 5 we briefly comment
about supersymmetric giant gravitons in the deformed background. In
Section 6 we explicitly demonstrate through examples and general
arguments that the results of Sections 3 and 4 
agrees with the field theory analysis which is performed in
details. Finally, in the Appendices we collect various technical
proofs, arguments and examples.

\section{$\beta$-deformation in toric theories}

\subsection{$\beta$-deformed quiver gauge theories}\label{betadef}

The entire class of super-conformal gauge theories living on D3-branes at 
toric conical Calabi-Yau singularities admits marginal deformations.
The most famous example is the $\beta$-deformation of ${\cal N}=4$ SYM
with $SU(N)$ gauge group where the original superpotential 
\beq
\label{N=4} \Phi_1 \Phi_2 \Phi_3 - \Phi_1 \Phi_3 \Phi_2 
\eeq
is replaced by the $\beta$-deformed one
\beq
\label{N=4beta} 
e^{i\pi \beta}\Phi_1 \Phi_2 \Phi_3 - e^{-i\pi \beta}\Phi_1 \Phi_3
\Phi_2 \, .  
\eeq 
A familiar argument due to Leigh and Strassler
\cite{LS} shows that the $\beta$-deformed theory is conformal for all
values of the $\beta$ parameter.

Similarly, a $\beta$-deformation can be defined for the conifold
theory. The gauge theory has gauge group $SU(N)\times SU(N)$ and
bi-fundamental fields $(A_i)_\alpha^A$ and $(B_p)_A^\alpha$ with
$\alpha,A=1,...,N,i,p=1,2$ transforming in the representations $(2,1)$ and $(1,2)$
of the global symmetry group $SU(2)\times SU(2)$,
respectively, and superpotential 
\beq\label{conifold} 
A_1 B_1 A_2 B_2 - A_1 B_2 A_2 B_1 \, .  
\eeq 
The $\beta$-deformation corresponds to the
marginal deformation where the superpotential is replaced by
\beq\label{conifoldbeta} e^{i\pi \beta} A_1 B_1 A_2 B_2 - e^{-i\pi
  \beta} A_1 B_2 A_2 B_1 \, .  \eeq

Both theories discussed above possess a $U(1)^3$ geometric symmetry corresponding to the isometries of the
internal space, one $U(1)$ is an R-symmetry while the other two act on the fields as flavour global 
symmetries\footnote{This $U(1)^3$ symmetry can be enhanced to a non abelian
 one in special cases. For instance it is $SU(4)$ for ${\cal
    N}=4$ SYM and $SU(2)\times SU(2)\times U(1)_R$ for the conifold.  In
  addition the conifold possesses a $U(1)_B$ baryonic symmetry. A
  generic toric quiver, besides the geometric symmetry $U(1)^3 = U(1)^2_F \times  U(1)_R $,
  presents several baryonic $U(1)$ symmetries. In this paper we will only be interested in the geometric symmetries 
of these theories.}.
The $\beta$-deformation is strongly related to the existence of such
$U(1)^3$ symmetry  and has a nice and
useful interpretation in terms of non-commutativity in the internal
space \cite{LM}.  The deformation is obtained by selecting in $U(1)^3$ the two flavour
symmetries $Q_i$  commuting with the supersymmetry charges
and using them to define a modified non-commutative product.  This
corresponds in field theory to replacing the standard product between
two matrix-valued elementary fields $f$ and $g$ by the star-product
\begin{equation}
\label{star}
f*g \equiv e^{i \pi \beta (Q^f \wedge Q^g)} f g 
\end{equation}
where $Q^f=(Q^f_1,Q^f_2)$ and $Q^g=(Q^g_1,Q^g_2)$ are the charges of the matter fields
under the two $U(1)$ flavour symmetries and
\beq
\label{wedge} 
(Q^f \wedge Q^g)= (Q^f_1 Q^g_2 -Q^f_2 Q^g_1) \, .
\eeq

The $\beta$-deformation preserves the $U(1)^3$ geometric symmetry 
of the original gauge theory, while
other marginal deformations in general further break it.

All the superconformal quiver theories obtained from toric Calabi-Yau
singularities have a $U(1)^3$ symmetry
corresponding to the isometries of the Calabi-Yau and therefore admit exactly marginal
$\beta$-deformations.  The theories have a gauge group $\prod_{i=1}^G
SU(N)$, bi-fundamental fields $X_i$ and a bipartite structure which is
inherited from the dimer construction \cite{dimers}.   The superpotential contains an
even number of terms $V$ naturally divided into $V/2$ terms weighted
by a $+1$ sign and $V/2$ terms weighted by a $-1$ sign \beq
\sum_{i=1}^{V/2} W_i(X) - \sum_{i=1}^{V/2} \tilde W_i(X) \, .  \eeq
The $\beta$-deformed superpotential is obtained by replacing the ordinary
product among fields with the star-product (\ref{star}) and, as
discussed in Appendix B, can always be written after rescaling fields as \cite{HB}
\beq\label{defSup} 
e^{i \alpha \pi \beta} \sum_{i=1}^{V/2}
W_i(\varphi) - e^{-i \alpha \pi \beta} \sum_{i=1}^{V/2} W_i(\varphi)
\eeq 
where $\alpha$ is some rational number.  It is obvious how ${\cal
  N}=4$ SYM and the conifold fit in this picture; other examples will
be given in Section \ref{gaugetheory}.

\bigskip
 
The $\beta$-deformation drastically reduces the mesonic moduli space of the theory,
which is originally isomorphic to the $N$-fold symmetric product of
the internal Calabi-Yau. To see quickly what happens consider the case
where the $SU(N)$ groups are replaced by $U(1)$'s - by abuse of language
we can refer to this as the $N=1$ case. Physically, we are considering
a mesonic direction in the moduli space where a single D3-brane is
moved away from the singularity. In the undeformed theory the D3-brane
probes the Calabi-Yau while in the $\beta$-deformed theory it can only
probe a subvariety consisting of complex lines intersecting at the
origin. This can be easily seen in ${\cal N}=4$ and in the conifold
case.

For ${\cal N}=4$ SYM the
F-term equations read
\beq
\label{N4}
\Phi_i \Phi_j = b\, \Phi_j \Phi_i, \qquad (i,j) =(1,2),(2,3)\, {\rm
  or}\, (3,1) \eeq where $b=e^{-2i \pi\beta}$. Since $\Phi_i$ are
c-numbers in the $N=1$ case, these equations are trivially satisfied
for $\beta=0$, implying that the moduli space is given by three
unconstrained complex numbers $\Phi_i$, giving a copy of
$\mathbb{C}^3$.  However, for $\beta\ne 0$ these equations can be
satisfied only on the three lines given by the equations
$\Phi_j=\Phi_k=0$ for $j\neq k$. Only one field $\Phi_i$ is different
from zero at a time.

For the conifold the F-term equations read
\bea\label{conFterm}
B_1 A_1 B_2 &=&  b^{-1}\, B_2 A_1 B_1 \, ,\nonumber\\     
B_1 A_2 B_2 &=&  \,b\,\,  B_2 A_2 B_1 \, ,\nonumber\\  
A_1 B_1 A_2 &=&  \,b\,\,  A_2 B_1 A_1 \, ,\nonumber\\  
A_1 B_2 A_2 &=&  b^{-1}  A_2 B_2 A_1 \, .
\eea
These equations are again trivial for $\beta=0$ and $N=1$, 
the fields becoming commuting c-numbers.
The  brane moduli space is parametrized by the four gauge invariant mesons
\beq x=A_1B_1,\,\,\,\, y=A_2B_2,\,\,\,\, z=A_1B_2,\,\,\,\,  w=A_2B_1\eeq
 which are not independent but
subject to the obvious relation $xy=zw$. This is the familiar description
of the conifold as a quadric in $\mathbb{C}^4$. For
$\beta\neq 0$, the F-term constraints (\ref{conFterm}) 
are solved when exactly one field $A$ and 
one field $B$ are different from zero. This implies that only one meson
can be different from zero at a time. The moduli space thus reduces to the
four lines 
\beq
\label{conifoldlines}  y=z=w=0 \, , \qquad x=z=w=0 \, , 
\qquad x=y=z=0 \, , \qquad x=y=w=0 .
\eeq

We will see in Section \ref{dualgiant} using the dual gravity solutions
and in Section \ref{gaugetheory} using field theory that for all
$\beta$-deformed toric quivers the abelian 
mesonic moduli space is reduced to $d$
complex lines, where $d$ is the number of vertices in the toric diagram
of the singularity.

Something special happens for $\beta$ rational. New branches in the
moduli space open up. The ${\cal N}=4$ case was originally discussed
in \cite{bl} and the conifold in \cite{tatar}. In all cases these
branches can be interpreted as one or more branes moving on the
quotient of the original Calabi-Yau by a discrete $\mathbb{Z}_n\times
\mathbb{Z}_n$ symmetry. We will describe these branes explicitly in
the gravitational duals in Section \ref{dualgiant}. The field theory
analysis of these vacua requires a little bit of technical patience
and it is deferred to Section \ref{gaugetheory}.

\subsection{$\beta$-deformed toric manifolds}\label{CYgeom}

The general prescription for determining the supergravity dual of a
$\beta$-deformed theory has been given by Lunin and Maldacena
\cite{LM}. The original background has a $U(1)^3$
isometry and the prescription amounts to performing
a particular T-duality along two $U(1)$ directions commuting with
the supersymmetry charges.

For a quiver gauge theory, the undeformed gravity solution is a warped product of 
4-dimensional Minkowski times a Calabi-Yau cone over  a Sasaki-Einstein manifold
\beq
\label{met4plus6}
{\rm d}s^2_{10} = e^{2 A} {\rm d}s^2_4 + e^{- 2 A}  {\rm d}s^2_6 \, ,
\eeq
where the warp factor is $e^{2 A} = r^2$. In all the formulae we are omitting 
factors of the radius of Anti de Sitter (see footnote 3 at page 9).

In the toric case these Calabi-Yaus have exactly three isometries and
the Lunin--Maldacena method can be applied. In \cite{LM} the $\beta$-deformation 
of the conifold and of $Y^{pq}$ spaces are explicitly
computed using the known metrics for these Sasaki-Einstein spaces.  In
this paper we consider the general case of a toric Calabi-Yau cone. We
will show that, as usual, most computations regarding supersymmetric
quantities can be performed without knowing the explicit form of the
metric.  We will just need the general characterisations of the
Calabi-Yau metrics given in \cite{MSY1} which we now review.

\subsubsection{The geometry of toric Calabi-Yau cones}

\begin{figure}[ht]
\begin{center}
\includegraphics[scale=0.45]{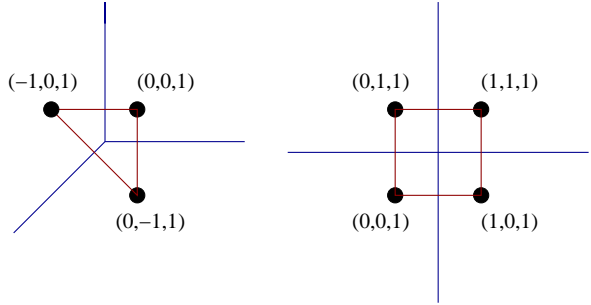} 
\caption{The toric diagram for $\mathbb{C}^3$ and the conifold
  consisting of the points $V_{\alpha}=(v_{\alpha},1)$ pictured in the plane $z=1$
  in $\mathbb{R}^3$. The vectors $V_{\alpha}$ determine a rational polyedron
  in $\mathbb{R}^3$ with three and four sides, respectively, whose
  projection on the plane $z=1$ is shown in the Figure.} 
\label{fibration}
\end{center}
\end{figure}

The geometry of a toric Calabi-Yau cone is
completely determined by $d$ integer vectors $V_{\alpha} \in \mathbb{Z}^3$. In fact 
there is a very explicit description of toric cones as $T^3$ fibrations over
a rational polyedron described by \cite{MSY1}
\beq
\label{cone}
{\cal C}^* = \{ {\rm y}\in \mathbb{R}^3 | l_{\alpha} ( {\rm y}) = V^{i}_{\alpha} y_i
\ge 0, \, \alpha = 1 \ldots d \} \eeq where $V_{\alpha}$ are the inward pointing
vectors orthogonal to the facets of the polyedral cone. The $T^3$
fibration degenerates to $T^2$ on the facets of the
polyedron, $l_{\alpha}({\rm y})=0$, and further degenerates to $T^1$ on the edges (intersections
of two facets). As a simple example, the trivial Calabi-Yau
$\mathbb{C}^3$ parametrized by three complex variables $Z_i= \sqrt{2
  y_i} e^{i\psi^i}$ can be considered as a $T^3$ fibration,
parameterised  by the three angles $\psi^i$, over the first octant in
$\mathbb{R}^3$ given by the three equations $y_i\ge 0$. Here
$V_1=(1,0,0)$, $V_2=(0,1,0)$, and $V_3=(0,0,1)$. In the following we will make a
convenient change of coordinates in order to have the third coordinate
of all $V_{\alpha}$ equal to one. Similarly, the conifold can be described as
a $T^3$ fibration over a polyedron with four sides, as shown in Figure
\ref{fibration}.

As shown in \cite{MSY1} the metric on the Calabi-Yau cone can be 
written as
\beq
\label{sympcoor} {\rm d}s^2_6 = g^{ij} {\rm d}y_i {\rm d}y_j +g_{ij} {\rm d}\phi^i{\rm d}\phi^j
\eeq 
with $g^{ij}$ the inverse matrix of $g_{ij}$.  Due to the toric
condition, $g_{ij}$ only depends on the variables $y_i$; the metric is
a cone if and only if $g^{ij}$ is homogeneous of degree $-1$ in $y$.
Regularity of the metric implies that near the facets 
\beq\label{reg}
g^{ij}= \sum_{\alpha=1}^d \frac{V^{i}_{\alpha} V^{j}_{\alpha}}{l_{\alpha}({\rm y})} + \mbox{regular
  terms} \, .  
\eeq 
The Calabi-Yau condition further requires that the
vectors $V_{\alpha}$ lie on a plane. We will choose coordinates where $V_{\alpha} =
(v_{\alpha},1)$.  The integer points in the plane, $v_{\alpha}$, describe the toric
diagram of the Calabi-Yau.

As in \cite{MSY1} we can also use complex coordinates to describe the manifold
\beq
\label{zcoor}
z^i = 
x^i + i \phi^i \, .
\eeq
A K{\"a}lher metric can be written in terms of a K{\"a}lher potential $F(z^i)$.
In the toric case $F$ only depends on the real part, $x^i$, of the coordinates
so that, if we define
\beq
\label{Fpot}
g_{ij} = \frac{\partial^2 F}{\partial x^i \partial x^j} \, ,
\eeq 
the metric can be written as
\beq
\label{CY}
{\rm d}s^2_6 = g_{ij} {\rm d}z^i {\rm d}\bar{z}^j = g_{ij} {\rm d}x^i {\rm d}x^j +g_{ij} {\rm d}\phi^i{\rm d}\phi^j \, .
\eeq
There is a nice relation between symplectic and complex coordinates given by
\beq\label{change}
y_i=\frac{\partial F}{\partial x^i}
\eeq
and, as the notation suggests, 
the function $g_{ij}({\rm x})$ appearing in the complex coordinates form 
of the metric is the same as the function $g_{ij}({\rm y})$ appearing in the 
symplectic form of the metric after changing variables from ${\rm x}$ to
${\rm y}$.   

The K{\"a}hler form and the holomorphic three-form are given by
\bea
\label{kahlercy}
J_{(0)} &\equiv& \frac{i}{2} g_{ij} {\rm d} z^i \wedge {\rm d} {\bar z}^j\, , \\
\label{omegacy}
\Omega_{(0)} &\equiv& e^{i\alpha} \sqrt{\det g_{ij}} {\rm d} z^1 \wedge  {\rm d} z^2 
\wedge {\rm d} z^3 \\
\label{exp}
&=&  e^{x^3 + i \phi^3} {\rm d} z^1 \wedge  {\rm d} z^2 
\wedge {\rm d} z^3 \, .
\eea
As shown in \cite{MSY1}, the explicit form of $\Omega_{(0)}$ given in (\ref{exp}) 
follows from Ricci-flatness, which implies $\det g_{ij}=e^{2x^3}$, and correlates
the phase in $\Omega_{(0)}$  with the complex direction $z^3$ associated with 
the third component of the vectors $V_{\alpha} = (v_{\alpha},1)$. 

The R-symmetry of the gauge theory is dual to the Reeb vector of the
Sasaki-Einstein space
\beq\label{reeb}
K=\sum_{i=1}^3 b^i \frac{\partial}{\partial\phi^i} \, ,
\eeq
where the components $b^i = 2 g^{ij} y_j$ turn out to be constants \cite{MSY1}. 
Moreover the third component $b_3$ is set to $3$ by the Calabi-Yau condition.
The vector $b=(b^i,3)$  satisfies
\beq\label{reeb2}
g_{ij} b^i b^j =r^2 .
\eeq

The Reeb vector $K$ is the partner under the complex structure of the dilatation
operator $r \partial_r$. Notice that the conical form of the metric is hidden
both in the symplectic and complex coordinates. The very same radial coordinate
$r$ is given by a non-trivial expression depending on the actual value of
the Reeb vector
\beq\label{rad}
r^2=2 b^i y_i \, .
\eeq 
 
\subsubsection{The $\beta$-deformed Calabi-Yau}
The $\beta$-deformation of toric Calabi-Yaus can be obtained as in \cite{LM}.
For simplicity we will consider $\beta$ real in the following.
We consider a two-torus in the internal manifold and we perform a T-duality 
transformation that acts on the complexified K{\"a}hler modulus of 
the two-torus as
\beq
\label{sl22}
\nu = B_{T^2} + i \sqrt{\det g_{T^2}} \rightarrow \frac{\nu}{1 +
  \gamma \nu} \, .  \eeq Here we choose the $T^2$ in the directions
$(\phi_1, \phi_2)$ since the action leaves the holomorphic three-form
invariant.  The parameter $\gamma$ in supergravity is proportional to
the $\beta$-parameter in the gauge theory.

The T-dual metric and B-field can be computed via  Buscher rules 
\beq
\label{buscher}
E = g - B_2 \rightarrow  (d E + c) (a E + b)^{-1}
\eeq
by embedding
the $O(2,2)$ transformation (\ref{sl22}) in $O(6,6)$
\beq
\label{OLM}
O_{LM} = \left( \begin{array}{cc} a & b\\
                                  c & d \end{array} \right) =
\left( \begin{array}{cc} Id_6 & \beta \\
                                  0 & Id_6 \end{array} \right) \, ,
\eeq
where the bivector $\beta$ is defined as
\beq 
\beta = \gamma \left( \begin{array}{ccc} 0_3 & 0 & 0 \\
                              0 & i \sigma_2 & 0 \\
                              0 & 0 & 0\end{array} \right) \, .
\eeq

The choice of the two-torus introduces a four plus two splitting in the metric
that can be made explicit by rewriting it in the following form
\beq
\label{metcy42}
{\rm d}s^2_6 = h_{ab} \chi^a_{(0)} \bar{\chi}^b_{(0)} +  Z \bar{Z} \qquad a,b=1,2
\eeq
%
where $h_{ab} = g_{ab}$ is the metric on the two-torus
and we have defined the one-forms
\bea
\label{chis}
\chi^a_{(0)} &=& {\rm d}z^a +  h^{ac} g_{c3} {\rm d}z^3  \qquad a=\, 1,2 \, ,\\
& = & ({\rm d}x^a +  h^{ac} g_{c3} {\rm d}x^3) 
+ i ({\rm d}\phi^a +  h^{ac} g_{c3} {\rm d}\phi^3) 
= X^a + i Y^a \\
\label{zeta}
Z  &=& e^{i\phi^3}\sqrt{g_{33} -  h^{ab} g_{a3} g_{b3}} \, {\rm d}z^3 
= \frac{{\rm d} w^3}{r^2 \sqrt{h}}
\eea
with $h = \det(h_{ab})/r^4$. The subscript $(0)$ is to distinguish these forms
from the corresponding one in the T-dual background. 
We also defined $w_3=e^{z^3}$.
The one form $Z$ parameterises the direction orthogonal to the two-torus
and to pass from the first to the second expression in (\ref{zeta}) we used the 
identity
\beq
\label{detg}
\det(g_{ij}) = e^{2 x^3} = \det(h_{ab}) (g_{33} - h^{ab}  g_{a3} g_{b3} ) \, .
\eeq

The advantage of writing the metric as in (\ref{metcy42}) is that the
T-duality transformation (\ref{OLM}) results simply in a rescaling of its
angular part 
\beq
\label{metbeta42}
{\rm d}s^2_6 = h_{ab} X^a X^b 
+ G \, h_{ab} Y^a Y^b +  Z \bar{Z}
\eeq
by the function
\beq
\label{Gtoric}
G = \frac{1}{1 + \gamma^2 h} \, .
\eeq

The antisymmetric part of (\ref{buscher}) gives the  NS two-form of 
the $\beta$-deformed solution
\beq
\label{B2beta}
B  = \gamma \, h \, G \,  Y^1 \wedge Y^2 \, .
\eeq
The dilaton and the warp factor are
\beq
\label{dilA}
e^{\Phi}=\sqrt{G} \, , \qquad e^A = r \, ,
\eeq 
respectively, while the
non-vanishing RR fields are given by\footnote{In all the formulae for
  the background we are understanding factors of the $AdS_5$ radius,
  $L$, which is given by: $L^4=4 \pi^4 g_s N \alpha'^2 / Vol(X_5)$,
  where $N$ is the number of $D3$-branes and $X_5$ is the undeformed
  Sasaki-Einstein manifold. In particular the metric $ds^2_{10}$ has a
  factor of $L^2$, the NS flux $H$ a factor of $L^4$, $F_3$ and
  $F_5$ a factor of $L^4/g_s$ and $G$ should be defined as: $G^{-1}=1
  + \gamma^2 L^4 h$. Our formulae are in the  string frame and we will set
  $\alpha'=1$.}
\bea
\label{F5d}
&& F_{5} = 4 {\rm vol}_4 \wedge \frac{{\rm d}r}{r} 
+ 4 G {\rm vol}_{X_5} \, , \\
\label{F3d}
&&  F_3  = - 4 \gamma \, \omega_2 \wedge {\rm d}  \phi^3 = {\rm d} C_2 \, ,
\eea
where ${\rm vol}_{X_5} = \ast_6 \frac{{\rm d}r}{r} = \omega_2 \wedge {\rm d} \phi^1 \wedge {\rm d} \phi^2 \wedge 
{\rm d} \phi^3$ is the volume form of the undeformed Sasaki-Einstein manifold $X_5$, and the closed form $\omega_2$ depends only on the $x^i$ coordinates.

\subsection{The $\beta$-deformed pure spinors}\label{pure}

Recently it has been shown that a unifying formalism to treat ${\cal N}=1$
compactifications with non trivial background fluxes is provided by 
Generalised Complex Geometry. For a detailed discussion
of pure spinors, Generalised Complex Geometry and its applications to
string theory see \cite{Hitchin,gualtieri,GMPT3}; here we will very briefly summarise
what we will need in the following section.

The idea is, given a manifold, to study objects
defined on the sum of the tangent and cotangent bundles, $T \oplus T^*$.
We can for instance define spinors on $T \oplus T^*$: these will 
be SO(6,6) spinors and have a representation in terms of differential 
forms of mixed degree, $\Lambda^{\bullet}(T^*)$. We call pure the spinors that are
annihilated by half of the generators of Cliff(6,6).  They are represented by  sum
of even and odd forms, $\Phi_\pm$, corresponding to
the positive and negative chirality, respectively.

The relevance for supergravity lies in the observation that such pure spinors
can be  obtained as tensor products
of ordinary spinors. More precisely, if 
we decompose the type IIB ten-dimensional 
supersymmetry parameters as 
\beq
\varepsilon^i = \zeta_+ \otimes \eta^i_+ + \zeta_- \otimes \eta^i_- \, ,
\eeq
where $\zeta_+$ ($\zeta_- = \zeta_+^*$) and $\eta^i_+$ 
($\eta^i_- = \eta_+^{i *}$) 
are positive chirality spinors in four and six dimensions, the pure spinors are defined as
\bea
\label{purespinorsdef}
&& \Phi_+ = \eta^1_+ \otimes \eta^{2 \dagger}_+ \, ,\\
&& \Phi_- = \eta^1_+ \otimes \eta^{2 \dagger}_-  \, .
\eea
The spinors constructed this way define an SU(3) $\times$ SU(3) structure on
$T \oplus T^*$ \footnote{The pure spinors must obey the SU(3)$\times$ SU(3) 
compatibility 
conditions $\langle \Phi_-,{\cal X}\cdot \Phi_+ \rangle
= \langle  \Phi_-,{\cal X}\cdot \bar{\Phi}_+ \rangle =0$ for any element ${\cal X} = X + \xi$ of
$T \oplus T^*$, where $X$ and $\xi$ are a vector and a one-form, respectively.}. 
By introducing an inner product between forms (Mukai pairing)
\beq
\label{mukai}
\langle A,B \rangle \equiv (A \wedge \lambda(B) ) |_{\rm top} \qquad \qquad  
\lambda(A_n) = (-)^{{\rm Int}[n/2]} \, ,
\eeq
we can define the norm of the pure spinors as
\beq
\label{norms}
\langle \Phi_+,\bar{\Phi}_+ \rangle = \langle  \Phi_-,\bar{\Phi}_- \rangle 
= -\frac{i}{8} ||\Phi||^2 \, {\rm vol}_6 
=   -\frac{i}{8} ||\eta_1||^2 ||\eta_2||^2 \, {\rm vol}_6  \, .
\eeq

It is  convenient to introduce normalised twisted spinors
\beq
\label{twistedspinors}
\hat{\Psi}_{\pm} = e^{-\Phi} e^{-B} \wedge \Psi_{\pm} = \frac{8
  i}{||\Phi||} e^{-\Phi} e^{-B} \wedge \Phi_{\pm} \, .  
\eeq 
All the NS content of the background (internal metric,
$B$ field and dilaton) can be extracted from
$\hat{\Psi}_\pm$. Moreover the twisted pure spinors are those
transforming nicely under T-duality.

Using the above definition as bispinors, it is possible to 
rewrite the  supersymmetry conditions
for type IIB supergravity as differential equations for the pure spinors $\hat{\Psi}_\pm$
\bea
\label{susyeqs}
&& {\rm d}( e^{3 A} \hat{\Psi}_-) = 0 \, ,\\
\label{susyeqs1}
&& {\rm d} (e^{2 A } \, {\rm Im} \hat{\Psi}_+) = 0 \, , \\
\label{susyeqs2}
&& {\rm d} (e^{4 A} \, {\rm Re}\hat{\Psi}_+) = - e^{4 A} e^{-B} \ast
\lambda( F) \, .  
\eea 
Here the $\ast$ is with respect to the six
dimensional internal metric $e^{-2A} {\rm d}s^2_6$  and $F$
is the sum of the internal magnetic fields $F = F_1 + F_3 + F_5$. It
is related to the ten-dimensional RR fields as $F^{(10)} = F +
{\rm vol}_4 \wedge \lambda (\ast F)$.  The ten-dimensional
Bianchi identity $({\rm d} - H \wedge) F^{(10)}=0$ yields the Bianchi identity
 and the
equations of motion for $F$: $({\rm d}-H \wedge) F=0$ and $({\rm d}+ H\wedge)
(e^{4A} \ast F) =0$, respectively. Notice that the equations of motion
follow automatically from (\ref{susyeqs2}).

The pure spinor satisfying ${\rm d}(e^{3 A} \hat{\Psi}) = 0$, defines a twisted
generalised Calabi-Yau \cite{Hitchin, gualtieri}. Thus one can
interpret the closure of the pure spinor coming from the supersymmetry
variations as the generalisation to the flux case of the standard
Calabi-Yau condition for fluxless compactifications: all ${\cal N}=1$
vacua are Generalised Calabi-Yau manifolds \cite{gmpt2}.

\bigskip

The explicit form of the pure spinors depends on how the 
internal supersymmetry parameters $\eta^i$ are related to the globally defined
spinors on the manifold.
For the toric Calabi-Yau manifolds there is one globally defined (in this case
covariantly constant) spinor, $\eta_+$, so that one can choose
\beq
\eta^1_+ = e^{A/2} \eta_+,    \qquad  \eta^2_+ = i e^{A/2} \eta_+ \, ,
\eeq
and the pure spinors are given in terms
of the K{\"a}lher form and holomorphic three-form
\bea
\label{Psiminus0}
&& \hat{\Psi}^{(0)}_- = e^{-3 A} \Omega_{(0)} =   e^{- 3 A} {\rm d}z^1\wedge {\rm d}z^2\wedge {\rm d}w^3
\, , \\
\label{Psiplus0}
&& \hat{\Psi}^{(0)}_+ =  e^{- i e^{-2 A} J_{(0)}} =  e^{1/2 \, e^{-2 A} g_{ij} {\rm d}z^i\wedge {\rm d}\bar z^j} \, .
\eea
In the Calabi-Yau background the dilaton and the NS two-form are zero, so that there 
is no difference between twisted and untwisted spinors.  

\bigskip

We now want to construct the pure spinors corresponding to the $\beta$-deformed 
backgrounds as the T-duals of the Calabi-Yau ones. 
As shown in \cite{HH} the T-duality transformation (\ref{OLM}) on the pure spinors is given 
by 
\beq
\label{betatransf}
\hat{\Psi}^{(0)} \rightarrow  \hat{\Psi} = e^{\beta} \cdot \hat{\Psi}^{(0)} 
= (1 + {\beta}) \cdot \hat{\Psi}^{(0)} \, ,
\eeq
where ${\beta}$ is a bivector associated with the two $U(1)$ isometries,
$\phi^1$ and $\phi^2$, of the Calabi-Yau. It acts on the pure spinor by
contractions\footnote{A generator of $O(6,6)$ acts linearly on the elements of 
$T \oplus T^*$.  If we define a generic element of $T \oplus T^*$ 
as $(X, \xi)$, with $X$ a vector and $\xi$ a one form, we have
\beq
\label{Oddtransf}
\left(\begin{array}{c} 
X \\
\xi 
\end{array} \right)  \rightarrow 
\left(\begin{array}{cc} 
A & \beta \\
B &  -A^T 
\end{array} \right) \, \left(\begin{array}{c} 
X \\
\xi 
\end{array} \right) \, ,
\eeq
where $A$ is an SO(6) element, $A = A^n_m{\rm d}x^m \otimes 
\iota_{\partial_{x^n}}$, $B$ is a two-form $ B 
= \frac{1}{2} B_{mn}{\rm d}x^m \wedge {\rm d}x^n$, and
$\beta$ is a bivector $\beta = \frac{1}{2} \beta^{mn} \iota_{\partial_{x^m}} \wedge  
\iota_{\partial_{x^n}}$. Then $O(6,6)$ element corresponding to the $\beta$-deformation, 
(\ref{OLM}), is just the bivector and 
and thus 
acts as in (\ref{betaaction}) on a generic differential form.}
\beq
\label{betaaction}
\beta =  \gamma \, \iota_{\partial_{\phi^1}} \wedge \iota_{\partial_{\phi^2}} =
 \gamma \, \iota_{\partial_{\phi^1}}  \iota_{\partial_{\phi^2}} \, .
\eeq 

Applying (\ref{betaaction}) to (\ref{Psiplus0}) and (\ref{Psiminus0}) we obtain a new pair of
pure spinors (here we have undone the twist) 
\bea
\label{Psiminus} \Psi_- &=&  \gamma \sqrt{G} e^{- 3 A}  {\rm d} w^3 \wedge
e^{\frac{1}{\gamma} {\rm d}z^1\wedge {\rm d}z^2 + B } 
\, ,\\
\label{Psiplus}
\Psi_+ &=&  \sqrt{G}  e^{- i e^{-2 A} J_{(0)}  
- \gamma h  X^1 \wedge X^2 + B } \, ,
\eea
where $B = \gamma \,  h \,  G Y^1 \wedge Y^2$ is the 
NS two-form of the $\beta$-deformed
background\footnote{It is a straightforward computation to show that these pure spinors are
equivalent to the dielectric ones in \cite{mpz} 
\bea
\label{Psiminusplus}
&&\Psi_-  = (-\sin 2\phi e^{i(\alpha+\beta)} e^{-A} z ) \wedge 
e^{i \frac{{\rm Re} \omega}{\sin 2\phi e^{2A}} 
-\cot 2\phi \frac{{\rm Im}\omega}{e^{2A}} }   \, , \\
&&\Psi_+ =
\left ( \cos 2\phi - i e^{-2A} j - \frac{\cos 2\phi}{2} e^{-2A} j^2 +\sin 2\phi e^{-2A}{\rm Im} \omega \right ) e^{ \frac{z\bar z}{2 e^{2A}}}  \nonumber
\eea
with $\sin 2\phi = -\gamma \sqrt{h} \sqrt{G}$, $\cos 2\phi = \sqrt{G}$. 
The  SU(2) structure
\bea
\label{betaSU(2)}
j &=&  \frac{i}{2}(\chi^1\wedge\bar\chi^1+\chi^2\wedge\bar \chi^2 )\, , \\
\omega &=& i \sqrt{h} \chi^1 \wedge \chi^2 \, ,
\eea
is defined in terms of the vielbein adapted to the $\beta$-deformed metric (\ref{metbeta42})
\beq
\chi^i = X^i +i \sqrt{G} Y^i \, .
\eeq
As before, the analogous quantities with superscript $(0)$ refer to the original Calabi-Yau 
metric. 
}. The usual SU(3)$\times$ SU(3) compatibility conditions between $\hat{\Psi}_-$ and  $\hat{\Psi}_+$
continue to hold since the Mukai pairing is invariant under a general $SO(6,6)$ 
transformation.

The expression for the closed pure spinor, (\ref{Psiminus}),
has a nice interpretation in terms of the generalised Darboux theorem \cite{gualtieri}. 
The pure spinors (\ref{Psiminus}), (\ref{Psiplus}) are of type $(1,0)$ 
and determine a splitting into four coordinates 
of symplectic type and two of complex type. 
The closure condition $ {\rm d}( e^{3 A}\hat{\Psi}_-) = 0$ implies the existence of
symplectic-complex coordinates $(\xi^i,z), i=1,..,4$ with
\beq
e^{3 A-\Phi} \Psi_-=e^{i k_0 + \tilde B}\wedge {\rm d}z \, ,
\eeq    
where $k_0={\rm d}\xi^1\wedge {\rm d}\xi^2+{\rm d}\xi^3\wedge {\rm d}\xi^4$ is the natural 
symplectic form and $\tilde B$ is a potential for $H$, ${\rm d}\tilde B=H$ 
\cite{gualtieri}. 
The symplectic coordinates predicted by the theorem are easily identified
from equation (\ref{Psiminus}) 
\beq
\frac{1}{\gamma}{\rm d}z^1\wedge {\rm d}z^2 +B 
\equiv \frac{i}{\gamma}({\rm d}x^1 \wedge {\rm d} \phi^2 
- {\rm d} x^2 \wedge {\rm d} \phi^1) + \tilde B 
\eeq
with the real and imaginary parts of the original complex coordinates of 
the Calabi-Yau $(x^i,\phi^i)$; $\tilde B = B + \frac{1}{\gamma}
({\rm d}x^1 \wedge {\rm d} x^2 - {\rm d} \phi^1 \wedge {\rm d} \phi^2)$.
We see that, although the $\beta$-deformed manifold looks very complicated
and it is not even a complex manifold, the generalised geometry selects
coordinates that are trivially related to the original complex coordinates
of the Calabi-Yau. As a consequence, all questions about supersymmetric
and BPS quantities in the $\beta$-deformed background can be still analysed
in terms of the original complex coordinates. This is not completely
unexpected, since the $\beta$-deformed ${\cal N}=1$ gauge theory has a natural
complex structure for all values of $\beta$.

\bigskip

In terms of the pure spinors it is straightforward to check that the T-dual background
is still supersymmetric. If we assume that  $\phi^{1,2}$ are 
supersymmetry-preserving isometries, ${\cal L}_{\partial_{\phi^{1,2}}}\hat{\Psi}=0$, then 
${\cal L}_{\partial_{\phi^{1}}}(\iota_{\partial_{\phi^2}}\hat{\Psi})=0$ and 
\bea
{\rm d} ({\beta}\cdot\hat{\Psi})=\gamma {\rm d} (\iota_{\partial_{\phi^1}}\iota_{\partial_{\phi^2}}
\hat{\Psi})
=-\gamma\iota_{\partial_{\phi^1}} {\rm d} (\iota_{\partial_{\phi^2}}\hat{\Psi})
=\gamma\iota_{\partial_{\phi^1}}\iota_{\partial_{\phi^2}} {\rm d} \hat{\Psi}={\beta}\cdot{\rm d}\hat{\Psi} \, .
\eea
Thus for a  $\hat{\Psi}$ which is invariant along $\phi^1,\phi^2$
\beq
\label{mainbeta}
{\rm d} (e^\beta\cdot \hat{\Psi})=e^\beta\cdot {\rm d} \hat{\Psi} \ .
\eeq
Then from  (\ref{mainbeta}) it follows that the T-dual spinors
satisfy the supersymmetry conditions, (\ref{susyeqs})-(\ref{susyeqs2}), if the original ones do. The 
T-dualised RR fields can be computed from $e^{-B} \ast \lambda (F) = e^{\beta}\cdot e^{-B^{(0)}} \ast  \lambda(F^{(0)})$. 
For the $\beta$-deformation of the quiver theories, this gives in particular 
\bea
&& F_5 = \ast {\rm d} (4A) = G F_5^{(0)}  \, , \\
&& F_3 = \ast (B \wedge  \ast F_5) \, , \qquad F_1 = 0 \, .
\eea
One can check that these are the same as in (\ref{F5d}) and (\ref{F3d}) and  satisfy (\ref{susyeqs2}) with the pure spinor given by (\ref{Psiplus}).

\bigskip

Finally, it is also easy to verify that the topology of the $\beta$-transformed background is the 
same as that of the original one, which was assumed to be smooth. 
The only points 
where one can have topology changes are the edges of the symplectic cone ${\cal C}^*$,
where the circles defined by $\phi^{1,2}$ shrink to zero. These are precisely the points 
where the bivector $\beta$ vanishes. To see this we can use the definition of the 
toric manifold as a $T^3$ fibration over the symplectic cone ${\cal C}^*$ \cite{MSY1}.
On the  $\alpha$-th facet of the cone ${\cal C}^*$ a given combination of the three 
angles $\phi^i$ degenerates. 
The precise combination can be read from the corresponding vanishing vector
\beq
K_{\alpha}=\sum_{i=1}^3 V^{i}_{\alpha} \frac{\partial}{\partial\phi^i} =  
v^{1}_{\alpha} \frac{\partial}{\partial\phi^1} +  v^{2}_{\alpha} \frac{\partial}{\partial\phi^2} 
+  \frac{\partial}{\partial\phi^3}
\eeq
where $V_{\alpha}=(v_\alpha,1)$ is the vector orthogonal to the facet. Thus, on the $\alpha$-facet only one linear combination of the three angles $\phi^i$ degenerates. This
is not enough in general to make the bivector $\beta$ vanishing. 
On the other hand, consider the edge of ${\cal C}^*$ corresponding to the
intersection of the $\alpha$-th and $\alpha+1$-th facets; the
vector $K_{\alpha}-K_{\alpha+1} = (v_\alpha-v_{\alpha+1})^1\partial_{\phi^1}+
(v_\alpha-v_{\alpha+1})^2\partial_{\phi^2}$ also vanishes. 
Since the (two-dimensional) integer vectors
$v^a_{\alpha}$ and $v^a_{\alpha+1}$ are not equal \footnote{Recall that 
$v_\alpha$ determines the toric diagram
of the Calabi-Yau so no consecutive $v_\alpha$ can be equal.}, it follows that the killing 
vectors $\partial_{\phi^1}$  and $\partial_{\phi^2}$ are proportional and $\beta$ vanishes. Thus $\beta$ vanishes precisely on the edges of the cone. 

If the original   
$SO(6,6)$  spinor $\hat{\Psi}^{(0)}$ is regular, then at these points
\bea
\beta\cdot \hat{\Psi}^{(0)} \rightarrow 0\ .
\eea
Thus, at these degenerate points 
\bea
\hat{\Psi} \simeq \hat{\Psi}^{(0)} \, .
\eea
Since a background is completely specified by $\hat{\Psi}_-$, $\hat{\Psi}_+$ and
$F$,  at the degeneration points the new background looks 
similar to the original one. Hence it is regular as well, as discussed from the
metric point of view in \cite{LM}.

\bigskip

\section{D3 and D5 probes}
The connection between gravity and field theory is provided by the
study of supersymmetric D-brane probes moving on the $\beta$-deformed
background.  We first analyse space-time filling static D-brane
probes, easily extending the results of \cite{LM} to a generic
Calabi-Yau background. A parallel analysis is performed for non-static
probes, in particular dual giant gravitons \cite{dual}, corresponding to brane
probes wrapping a three-sphere in $AdS_5$ and spinning in the internal
manifold. The case of dual giants in the $\beta$-deformed ${\cal N}=4$
SYM has been analysed in \cite{IN}.

In this Section we perform an analysis based on the effective Lagrangian
on the world-volume of a probe moving in the deformed background. In
the next Section we will discuss the same results from the point of view
of T-duality and supersymmetry, using the Generalised Geometry perspective.

\subsection{Static probes}\label{static}

The moduli space of space-time filling supersymmetric static four-branes
should reproduce the mesonic moduli space of the dual gauge theory.
In the undeformed background we just have a single type of static
supersymmetric probe, a D3-brane which can live at every point of the
internal manifold.  Correspondingly, the abelian moduli space of the
dual field theory is isomorphic to the Calabi-Yau cone. In the
deformed background, we have two different types of static
supersymmetric probes, D3-branes, and dielectric D5-branes wrapped on the
(T-duality) two-torus and stabilized by a world-volume flux \cite{LM}.
Supersymmetric D3-probes can only live on a set of intersecting
complex lines inside the deformed Calabi-Yau, corresponding to the
locus where the $T^3$ toric fibration degenerates to $T^1$.  This is
in agreement with the abelian moduli space of the $\beta$-deformed gauge
theory which indeed consists of a set of lines.  In the case of
rational $\beta$, there exist supersymmetric D5-probes with a moduli
space isomorphic to the original Calabi-Yau divided by a
$\mathbb{Z}_n\times \mathbb{Z}_n$ discrete symmetry.  This statement
is the gravity counterpart of the fact that for rational $\beta$ new
branches are opening up in the moduli space of the gauge theory
\cite{bl,dorey}.

\subsubsection{Static D3 probes}
Consider a static space-time filling D3-brane probe.
The dynamics is governed by the brane world-volume action
\beq
\label{d3wvaction}
S_{D3} = S_{BI} + S_{CS} = - T_3 \int {\rm d}^4 \zeta e^{-\Phi}\sqrt{- \det G_{\mu \nu}}
+ T_3 \int C_4 \, .
\eeq
$G_{\mu \nu}$ is the pull back of the space-time metric $g_{MN}$ to 
the world-volume of the D3-brane
\beq
\label{wvmetric}
G_{\mu \nu} = \frac{\partial X^M \partial X^N}
{\partial \zeta^{\mu}  \partial \zeta^{\nu}} g_{MN} \, ,
\eeq
where $(\zeta^0, \zeta^1, \zeta^2, \zeta^3)$ are the world-volume 
coordinates on the brane. The ten-dimensional metric is given by
\beq
{\rm d}s^2_{10} =
r^2 {\rm d}x_\mu {\rm d}x^\mu  +\frac{1}{r^2} {\rm d}s^2_{X_6} \, .
\eeq
By inserting in the BI and CS terms the explicit expression
of the background fields (\ref{dilA})-(\ref{F5d}), we see that a D3-probe feels a
potential given by
\beq
\int {\rm d}^4\zeta V(y_i) \sim \int {\rm d}^4\zeta\,  r^4 \left(\frac{1}{\sqrt{G}}-1\right ) \, ,
\eeq
where $y_i$ are the coordinates on the internal space.
The potential is positive definite and  vanishes when $G\equiv 1$ or 
equivalently $h\equiv 0$.
$h$ vanishes precisely along the edges of the cone ${\cal C}^*$,  
where the $T^3$ fibration degenerates to $T^1$. 
In fact, it is easy to see from the explicit behaviour of the metric
near the facets, given in equation (\ref{reg}), that $h$ is regular and 
non vanishing in the interior of the cone and also in the interior of the
facets. On the other hand, as follows from equation (\ref{reg}), 
on the edge where the adjacent facets $\alpha$ and $\alpha+1$ intersect, $h$ vanishes as
\beq
\label{edges} 
h\sim \frac{l_{\alpha}(y)l_{\alpha+1}(y)}{|<V_{\alpha},V_{\alpha+1}>|^2} \, .
\eeq
We conclude that a supersymmetric D3-probe can only move along the $d$ edges
of the symplectic cone. Recall that the topology of the deformed theory
is the same as that of the original Calabi-Yau, allowing to reason in terms
of fibrations. Moreover, locally, the metric near the degeneration locus
is substantially identical to the original one.    

We expect that a single D3-brane probes the
abelian moduli space of the dual gauge theory.
What we found is 
compatible with the results for ${\cal N}=4$ SYM and the conifold
discussed in Section \ref{betadef}. There we found that the abelian
moduli space consists of three and four lines, respectively. 
These lines exactly correspond to the edges of the polyedral cone
discussed in Section \ref{CYgeom}. From the gravity analysis we thus get the
general prediction that the abelian moduli space of toric quiver gauge 
theories is given by a collection of $d$ lines, where $d$ is the number of
external vertices of the toric diagram. We will verify explicitly this
prediction in Section \ref{gaugetheory} with field theory methods.

\subsubsection{Static D5 probes}
As noticed in \cite{LM} a D5-brane wrapped on the two-torus $(\phi^1,
\phi^2)$ with a world-volume flux ${\rm F}={\rm d}\phi^1\wedge {\rm
  d}\phi^2/\gamma$ is supersymmetric.  It is easy to see that a
similar configuration exists for all Calabi-Yau backgrounds. The
supersymmetric D5-brane can live at an arbitrary point in ($y_i,
\phi^3$) and can have additional moduli corresponding to Wilson lines
on the two-torus. It is interesting to analyse the moduli space of
such configuration, since it corresponds to a particular non abelian
branch of the dual gauge theory.

Consider therefore a D5-brane wrapping the two-torus spanned 
by $(\phi^1, \phi^2)$ in the internal manifold.  The corresponding embedding is 
\bea
\label{d5embedding1}
&& x^\mu = \zeta^\mu,\quad \phi^1 = \zeta^4, \quad \phi^2 = \zeta^5, \nn\\
&& \phi^3 = \phi^3(\zeta^\mu), \quad  y_i = y_i(\zeta^\mu) 
\qquad \mu = 0,1,2,3 \, ,
\eea
where we call $(\zeta^0, \ldots, \zeta^5)$ the 
world-volume coordinates on the brane. 
The world-volume action for a D5-brane is
\bea
\label{d5wvaction1}
S_{D5} & = & -T_5 \int {\rm d}^6 \zeta e^{- \Phi}
\sqrt{- \det (G - B + {\rm F})_{\alpha \beta}} \nn\\
& & + T_5 \int C_6 + C_4 \wedge ({\rm F} - B) + C_2 \wedge ({\rm F} - B) \wedge ({\rm F}
- B) \, ,
\eea 
where we define ${\rm F}=2 \pi \alpha' \mathcal{F}$, with
$\mathcal{F}$ dimensionless. We will set $\alpha'=1$ as in the other
supergravity computations.

For the six-dimensional metric we will use the expression (\ref{metbeta42})
in  symplectic coordinates 
\bea\label{metsymp}
{\rm d}s^2_{X_6} &=& g^{i j} {\rm d}y_i {\rm d}y_j  
+ \tilde{g}_{i j} {\rm d} \phi^i {\rm d} \phi^j \\ 
&=&  g^{i j} {\rm d}y_i {\rm d}y_j 
+ G h_{a b} {\rm d} \phi^a {\rm d} \phi^b +2 G g_{a3} {\rm d}\phi^a {\rm d}\phi^3
+ [g_{3  3} - (1 - G) h^{a b} g_{a 3} g_{b 3}] ({\rm d} \phi^3 )^2 \, . \nonumber
\eea
Here and in the rest of this section the indices $i,j$ and $a,b$ are summed 
over $1,2,3$ and $1,2$, respectively. All the functions in the above ansatz
depend on the coordinates $y_i$ only since the angular directions are isometries
of the background.

The pulled-back metric is given by
\beq
\label{d5pbmetric1}
\left( \begin{array}{ccc}
r^2\eta_{\mu\nu} +\frac{1}{r^2}\left ( g^{i j} \partial_\mu y_i 
\partial_\nu y_j + \tilde{g}_{3 3} \partial_\mu \phi^3 \partial_\nu \phi^3 \right )
& G \, \partial_\nu\phi^3 g_{1 3} 
& G \partial_\nu\phi^3 g_{2 3} \\
G \,  \partial_\mu\phi^3 g_{1 3}   & G \, h_{1 1} &  G \, h_{1 2}  \\
G \, \partial_\mu\phi^3 g_{2 3}   & G \, h_{2 1} &  G \, h_{2 2}
\end{array} \right) \, .
\eeq
Similarly the pull back of the B-field has components
\bea
\label{d5pbBfield1}
&& B_{\mu 4} = - \gamma \, h \, G (h^{2 a} g_{a 3}) \partial_\mu\phi^3  \, ,
\\
&& B_{\mu 5} =  \gamma \, h \, G (h^{1 a} g_{a 3})  \partial_\mu\phi^3   \, ,
\\
&& B_{45} =  \gamma \, h \, G \, .
\eea
The world-volume field strength has both magnetic and electric components
\beq
\label{d5fs1}
{\rm F}_{45} = \frac{1}{\gamma} \, , \qquad {\rm F}_{\mu 4} = \partial_\mu A_1(\zeta^\nu) \, ,
  \qquad {\rm F}_{\mu 5} = 
\partial_\mu A_2(\zeta^\nu) \, .
\eeq
The magnetic component is required by supersimmetry, while the electric
components correspond to space-time fluctuations of the Wilson lines on the 
two-torus.

Using the above expressions the determinant in the Born-Infeld action can be written as 
\beq
\label{d5detf1}
\det(G - B + {\rm F}) = r^6 \frac{G}{\gamma^2} \left [ \frac{1}{r^2}\left (g^{i  j} \partial_\mu y_i 
    \partial_\mu y_j + g_{3 3} (\partial_\mu\phi^3)^2 - 2 \gamma
    g_{3  a } \partial_\mu \phi^3 \hat f_\mu^a + \gamma^2 h_{a b}
    \hat f_\mu^a \hat f_\mu^b\right ) - r^2\right ] \, .  
\eeq 
where $\hat f_\mu^a= \epsilon^{ab}\partial_\mu A_b = \epsilon^{ab} {\rm F}_{\mu
  b}$.  The overall factor of $G$ cancels the contribution from the
dilaton so that the BI action for the D5-probe takes the
form\footnote{$S_{BI}$ and $S_{WZ}$ are proportional to $T_5 L^4
  \alpha' Vol(T^2)= \pi^2 N/(2 Vol(X_5))$. Not to clutter formulae we
  will only write a factor of $N$.}  
\beq
\label{d5BIstatic1}
S_{BI} = - \frac{N}{\gamma} \int {\rm d}^4\zeta r^3 
\sqrt{r^2 - \frac{1}{r^2}\left (g^{i  j} \partial_\mu y_i 
\partial_\mu y_j + g_{3  3} (\partial_\mu\phi^3)^2 - 2 \gamma  g_{3  a }  \partial_\mu \phi^3 \hat f_\mu^a +
\gamma^2 h_{a  b}  \hat f_\mu^a \hat f_\mu^b\right )} \, .
\eeq

The Wess-Zumino part of the action simplifies as well, since, as
noticed in \cite{LM}, the $C_6$ contribution cancels with $B_2 \wedge
C_4$.  The only non trivial contribution is 
\beq
\label{d5WZ1}
S_{WZ} = T_5 \int C_4 \wedge {\rm F}_{45} = \frac{N}{\gamma} \int {\rm d}t \, r^4 \, .
\eeq

The contribution to the potential vanishes for all values of the
moduli $y_i,\phi^3,A_a$. We then obtain a six-dimensional 
family of supersymmetric four-branes.

We want to discuss in detail the existence and the moduli space of
such configurations. First of all, due to charge quantisation, the
D5-brane solutions we find exist only for rational values of
$\gamma\equiv m/n$, as discussed in details in \cite{LM}\footnote{In \cite{LM} to see this 
they check that a configuration of
  $(N_{D3},N_{D5},N_{NS5})$ in the undeformed geometry is mapped to
  $(N_{D3},N_{D5}+ \gamma N_{D3},N_{NS5})$ by the Lunin-Maldacena
  transformation. Hence $\gamma=m/n$ and $N_{D3}=N$ must be a multiple
  of $n$.}. In fact, since the internal $T^2$ wrapped by the D5-brane
supports a flux ${\rm F}_{45}=1/\gamma$, there is an induced D3-charge that
has to be quantized.  If we set $\gamma=m/n$, with $m$ and $n$
relatively prime integers, we obtain a consistent configuration by
taking a D5-brane wrapped $m$ times on the contractible $T^2$
\footnote{In the case $m=1$ we can equivalently impose that the first
  Chern number for the $U(1)$ gauge bundle is integer: $\frac{1}{2\pi}
  \int_{T^2} \mathcal{F} = n$, which gives $\gamma=1/n$.}. This
configuration can be alternatively seen as a set of $n$ blown up
D3-branes.


Our solutions should correspond to additional branches of
the dual gauge theory which exist only for rational $\beta$. These are
well known for ${\cal N}=4$ SYM \cite{bl,dorey} and are discussed  in \cite{tatar} for
the conifold.  For a generic $\beta$-deformed quiver gauge theory we can study the geometry 
of these new branches by looking at the moduli space of the solutions.  For
simplicity consider the case $N_{D5}=m=1$.  
The moduli space of the brane is parameterised by $\{\phi^3, \tilde{\mathcal{A}}_a , y_i
\}$. $\phi^3$ and $y_i$, ($i=1,2,3$) are  four scalars
deformations corresponding to transverse movements of the D5-brane in the internal
geometry. Then we have two Wilson lines in the internal $T^2$,  corresponding to the 
deformations of the gauge field on the brane: $e^{i \int_a {\mathcal{A}}}$.
Here $\mathcal{A} = A/(2 \pi)$ such that ${\rm F}= {\rm d}A$, $\mathcal{F}= {\rm d} \mathcal{A}$ 
and the integral is over the two non trivial one cycles on $T^2$.
Notice that before 
T-duality the Wilson lines correspond to the position of the D3-brane on $T^2$.
Naively the space of the deformations of the gauge field is given by
the first cohomology of $T^2$, which is parametrized by  the
gauge invariants $\tilde{\mathcal{A}}_a= \int_{a} \mathcal{A}$, but since the holonomies, ${\rm exp}(i
\tilde{\mathcal{A}}_a)$, are the only physical observables, it is clear that they
have compact range: $0 \leq \tilde{\mathcal{A}}_a \leq 2 \pi$.

The metric for the moduli space 
can be read from the DBI action, when we give a space-time
dependence to all moduli. We can then interpret the electric field
strengths as the space-time derivatives of the Wilson lines: ${\rm F}_{\mu
  a}= \partial_\mu A_a = 2\pi \partial_\mu \mathcal{A}_a
\simeq \partial_\mu \int_{a} \mathcal{A}=\partial_\mu
\tilde{\mathcal{A}}_a $. By expanding (\ref{d5BIstatic1}) we obtain the
metric on the moduli space 
\beq S_{D5} =\frac{N}{2\gamma}\int {\rm
  d}^4 \zeta \left (g^{i \, j} \partial_\mu y_i
\partial_\mu y_j + g_{3 \, 3} (\partial_\mu\phi^3)^2 - 2 \gamma  g_{3 \, a }  \partial_\mu \phi^3 \hat f_\mu^a +
\gamma^2 h_{a \, b}  \hat f_\mu^a \hat f_\mu^b\right ) \, .
\eeq
This metric is identical to the metric of the original Calabi-Yau when we
identify 
\beq
\partial_\mu \phi^a= -\gamma \hat f_\mu^a \, , \qquad \mbox{or} \qquad
\phi^a \equiv -\gamma \epsilon^{ab} \tilde{\mathcal{A}}_b \, .
\eeq
As discussed above, for $m=1$ the angular variable $\phi^a$ associated to
the Wilson lines has period $2\pi/n$. We thus see that the metric on
the moduli space is just that of the original CY divided by
$\mathbb{Z}_n\times \mathbb{Z}_n$.

Therefore the prediction from the gravity analysis is that, for every
toric quiver gauge theory, at rational $\beta$, we have additional
Higgs branches isomorphic to the orbifold ${\rm CY} / \mathbb{Z}_n\times
\mathbb{Z}_n$. We will give evidence for this statement in Section
\ref{gaugetheory}.

\subsection{Dual giant gravitons}\label{dualgiant}



We are interested in this section in dual giant gravitons, brane
probes wrapping a three-sphere in global $AdS_5$ and spinning in the
internal manifold.  Dual giants are defined in global coordinates in
$AdS_5$. 



As shown in \cite{MSg}, the classical phase space of a supersymmetric
D3 dual giant on the undeformed Sasaki-Einstein background is isomorphic
to the original Calabi-Yau, that is the abelian moduli space of the dual
gauge theory. Upon geometric quantisation of the classical solutions one
obtains all the mesonic BPS states of the theory\footnote{By quantising the classical dual giant solutions we obtain
states of the gauge theory on $S^3\times \mathbb{R}$ \cite{dual}. All
these states are mapped to BPS operators via the conformal mapping to
$\mathbb{R}^4$.}.

In this section we will
extend this discussion and study the dynamics of the dual giant gravitons 
in the $\beta$-deformed geometries. 
Since the quantisation of the classical dual giant
solutions gives mesonic BPS states (corresponding to BPS operators),
we expect that the classical phase space of the dual giants contains
information about the mesonic moduli space of the dual gauge theory.
Dual giants for the $\beta$-deformed 
${\cal N}=4$ SYM were already analysed in \cite{IN}.


Exactly in parallel to the case of static probes, the $\beta$-deformed
geometries admit BPS dual giant gravitons of two kinds. The first type
of giants are present for all values of the deformation parameter
$\gamma$ and correspond to D3-branes wrapping an $S^3$ in $AdS_5$ and
spinning along the Reeb vector in the internal geometries. On the
field theory side they correspond to the operators parameterising the
abelian Coulomb branch of the theory.  The classical phase space of
the dual giants reproduces the abelian moduli space of the dual gauge
theory.  The other class of dual giants can exists only for rational
values of the deformation parameter and consists of D5-branes wrapping
the $S^3$ in $AdS_5$ and the two-torus $(\phi^1, \phi^2)$ in the
internal manifold. They rotate in the angular direction orthogonal to
the two-torus and have a magnetic world-volume field strength
proportional to $1/\gamma$.  The world-volume gauge field satisfies
the quantisation condition only for $\gamma$ rational. On the field
theory side these configurations correspond to Higgs branches that are
present when $\beta$ is rational.


\subsubsection{D3 dual giant gravitons}
\label{d3dualg}

We want to study the dynamics of a D3-brane probe that wraps the three-sphere
in $AdS_5$, written in global coordinates, and rotates on the internal manifold. 
This is still governed by the brane 
world-volume action (\ref{d3wvaction})
where we now take as ten-dimensional metric
\beq
{\rm d}s^2_{10} =
{\rm d}s^2_{AdS_5} + {\rm d}s^2_{X_5} \, .
\eeq 
The metric of $AdS_5$
is given in global coordinates \beq {\rm d}s^2_{AdS_5} = - V(R) {\rm
  d}t^2 + \frac{1}{V(R)} {\rm d}R^2 + R^2 ({\rm d} \theta^2 + \cos^2
\theta {\rm d} \alpha_1^2 + \sin^2 \theta {\rm d} \alpha_2^2 ) 
\eeq
with $ V(R) = 1 + R^2$. $t$ is the global time in $AdS_5$ and the
angles $\theta$, $\alpha_1 $ and $\alpha_2$ parameterise a round three-sphere. 
We will write the metric on $X_5$ as the restriction of the
six-dimensional internal metric to the hypersurface with $r=1$ 
\beq
\label{constraint}
2 b^i y_i =1 \, .
\eeq
>From now on, we consider as coordinates for $X_5$ the angles $\phi^i$
and two extra angles parameterised by the $y_i$ with the above constraint.

With this choice of coordinates the embedding $X^M (\zeta^{\mu})$
corresponding to the dual giant graviton can be taken as \bea
\label{d3embedding}
&& t = \tau, \quad R= R(\tau), \quad \theta = \zeta^1, \quad \alpha_1 = \zeta^2,
\quad \alpha_2 = \zeta^3 \, ,\nn\\
&& \phi^i = \phi^i(\tau), \quad y_i = y_i(\tau) \qquad i = 1, \dots, 3 \, .
\eea
It is then easy to see that
\beq
\label{d3detwvm}
\sqrt{- \det G_{\mu \nu}} = R^3 \cos \theta \sin \theta \Delta^{1/2} \, ,
\eeq
where we have defined (the dot represents the derivative with respect to
$t=\tau$) 
\beq
\label{delta}
\Delta = V(R) - \frac{\dot{R}^2}{V(R)} - g^{i  j} \dot{y}_i 
\dot{y}_j - \tilde g_{i  j} \dot{\phi}^i \dot{\phi}^j  \, .
\eeq
To evaluate the WZ term we can choose the pull back of the four-form
 potential to be
\beq
\label{4potential}
C_{(4)} = R^4 \sin \theta \cos \theta {\rm d}\tau  \wedge 
{\rm d}\theta \wedge   {\rm d}\alpha_1  \wedge {\rm d}\alpha_2 \, .
\eeq
Substituting (\ref{d3detwvm}) and (\ref{4potential}) into
(\ref{d3wvaction}) we obtain the Lagrangian for the
probe\footnote{Keeping into consideration also the factors of $L$, the
  Lagrangian for $D3$ dual giants is proportional to $T_3 L^4
  Vol(S^3)= \pi^3 N / Vol(X_5)$; however we will write explicitly only the factor $N$ in
  front of $\cal{L}$.}  
\beq
\label{lagrangian}
{\cal L} = - N R^3 (e^{- \Phi} \sqrt{\Delta} - R) \, .
\eeq

To find the explicit solutions for the possible motions of the D3-brane probe
it is convenient to pass to the Hamiltonian formalism and solve
the Hamilton equations of motion.
For the dual giant graviton we are considering the canonical momenta
are 
\bea
\label{d3momenta}
&& p_R = \frac{\partial \cal L}{\partial \dot{R}} = 
e^{- \Phi} \frac{N R^3}{\sqrt{\Delta}} \frac{\dot{R}}{V} \, ,\nn\\
&& p_{y_i} = \frac{\partial \cal L}{\partial \dot{y_i}} = 
e^{- \Phi} \frac{N R^3}{\sqrt{\Delta}} g^{i  j} \dot{y_j} \, ,\\
&& p_{\phi^i} = \frac{\partial \cal L}{\partial \dot{\phi^i}} = 
e^{- \Phi} \frac{N R^3}{\sqrt{\Delta}} {\tilde{g}_{i  j}} 
\dot{\phi^j} \, . \nn
\eea
The Hamiltonian then reads 
\bea
\label{hamiltonian}
{\cal H} &=& e^{- \Phi} \frac{N R^3}{\sqrt{\Delta}} V - N R^4 \nn\\
   &=& N R^3 (\sqrt{ V \Omega} - R) \, ,
\eea
where in the second line we have expressed everything in terms of the canonical
momenta and we have introduced the function 
\beq
\label{omega}
\Omega = e^{- 2 \Phi} + \frac{1}{N^2 R^6} ( V p_R^2 
+ g_{i j} p_{y_i} 
p_{y_j} + \tilde{g}^{i  j} p_{\phi^i} 
p_{\phi^j}) \, .
\eeq

The corresponding equations of motion are
\bea
\label{d3Ra}
&& \dot{R} = \frac{1 + R^2}{N R^2 x} p_R \, , \\
\label{d3Rb}
&& \dot{p}_R = N R^3 [4 - \frac{1}{x} (x^2 + 3 e^{- 2 \Phi}
+ \frac{(p_R)^2}{N^2 R^4} )]  \, ,\\
\label{d3ya}
&& \dot{y}_i = \frac{1}{N R^2 x}  g_{i  j} p_{y_j} \, ,\\ 
\label{d3yb}
&& \dot{p}_{y_i} = - \frac{N R^4}{2 x} \partial_{y_i} \Omega \, ,\\
\label{d3phia}
&& \dot{\phi}^i = \frac{1}{N R^2 x}  
\tilde{g}^{i  j} p_{\phi^j} \, ,\\
\label{d3phib}
&& \dot{p}_{\phi^i} = 0 \, ,
\eea
where we have defined
\beq
x = R \sqrt{\frac{\Omega}{V}} \, .
\eeq
A BPS solution representing a dual giant rotating in the internal manifold is
given by 
\bea
\label{d3soldil}
&& R = const \, , \qquad p_R = 0  \, ,\\
&& y_i = const \, ,  \qquad   p_{y_i} = 0  \, ,\\
&& \dot\phi^i = b^i \, , \qquad  p_{\phi^i} = 2 N R^2 y_i
\eea
with $y_i$ satisfying $\Phi(y_i)=0$.

To explicitly see it, it is convenient to introduce a set
of local angular coordinates adapted to the motion of the brane probe
\beq
\label{d3metricpsi}
ds^2_{X_5} = g^{i  j}  {\rm d}y_i  {\rm d}y_j+  H 
({\rm d} \psi + \sigma_a  {\rm d} \psi^a)^2 + h_{a  b} {\rm d}\psi^a 
{\rm d}\psi^b \, ,
\eeq
where $\psi$ is the angular direction in which the brane rotates, and the 
indices $a,b$ run from 1 to 2. As before the functions $H$ and $h_{ab}$ depend
on the variables $y_i$ only. In these coordinates the function $\Omega$ becomes
\beq
\label{omegapsi}
\Omega = e^{- 2 \Phi} + \frac{1}{N^2 R^6} ( V p_R^2 
+ g_{i  j} p_{y^i} 
p_{y_j} + H^{-1} p_{\psi}^2 + h^{a    b} (p_{\psi^a} - \sigma_a p_{\psi})
(p_{\psi^b} - \sigma_b p_{\psi})) \, ,
\eeq
while (\ref{d3phia}) and  (\ref{d3phib}) are substituted by
\bea
\label{d3psia}
& \dot{\psi} = \frac{1}{N R^2 x} ( H^{-1} p_{\psi} - h^{a  b} \sigma_a (p_{\psi_b}-\sigma_b p_{\psi}) ) \, ,  \quad
& \dot{p}_{\psi} = 0 \, ,\\
\label{d3ppsia}
& \dot{\psi}^a = \frac{1}{N R^2 x} h^{a \, b} (p_{\psi^b} - \sigma_b p_{\psi}) \, , \qquad 
\qquad \qquad \qquad & \dot{p}_{\psi^a} = 0 \, .
\eea

Since the brane rotates in the direction $\psi$ we expect
\beq
\label{condsol}
\dot{y}_i = 0, \qquad \dot{\psi}^a = 0, \qquad   \dot{R}=0 \, .
\eeq
The first condition, together  with (\ref{d3ya}) and  (\ref{d3yb}), implies
\beq
\label{condsoly}
p_{y_i} = 0 \qquad {\rm and} \qquad \partial_{y_i} \Omega = 0 \, .
\eeq
The second condition in (\ref{condsol}) imposes 
\beq
\label{condsolpsia}
p_{\psi^a} = \sigma_a p_{\psi} \, .
\eeq
And finally 
the third condition combined with (\ref{d3Ra}) and (\ref{d3Rb}) gives
\beq
\label{condsolR}
p_R =0 \qquad {\rm and} \qquad x = 2 \pm \sqrt{4 - 3 e^{-2 \Phi}} \, .
\eeq

Observe that the condition $\partial_{y_i}\Omega=0$ and the definitions of $x$
and $\Omega$ altogether imply
\beq \label{eqsmotion}
\partial_{y_i} \Phi =0, \qquad \partial_{y_i} H=0 \, .
\eeq

Up to now we have not imposed the condition that the dual giant must be 
BPS. This amounts to setting the Hamiltonian equal to the momentum in 
direction of the rotation
\beq
\label{BPScond}
{\cal H} = p_{\psi} \, .
\eeq

The value of $p_{\psi}$ and ${\cal H}$ on the solution are easily computed from the
equations above
\bea
\label{Hsol}
&& {\cal H} = N R^2 [x + R^2 (x -1)] \, ,\\
&& p_{\psi} = \sqrt{H} N R^2 \sqrt{R^2 (x^2 - e^{-2 \Phi}) + x^2}  \, ,
\eea
so that for the ratio to be equal to 1 for all values of $R$, one has to impose\footnote{There
might exist other solutions with fixed value of $R$. Most likely, an
analysis in terms of supersymmetry transformations would reveal that 
these solutions are not BPS. They would correspond
to truly 
isolated vacua in the dual field theory, that are not expected to exist
in such theories.}
\beq
\label{solother}
x = 1,  \qquad \Phi =0, \qquad H=1 \, ,
\eeq
which imply $\dot \psi=1$ on the BPS solutions.

We can now analyse the conditions for BPS motion. Let us start with
the case of the undeformed theory.  In the undeformed background,
$\Phi$ is identically zero. A supersymmetric configuration can be
obtained by allowing the probe to rotate along the Reeb vector.  In
fact the angle $\psi$ dual to Reeb vector is normalized to one 
\beq
H=g (K,K) = g_{ij} b^i b^j \equiv 1 \, , 
\eeq 
where we made use of
equation (\ref{reeb2}) on the Sasaki-Einstein $r=1$. Thus the BPS
equations (\ref{eqsmotion}) and (\ref{solother}) are satisfied.  This
reproduces the results found in \cite{MSg}: a supersymmetric dual
giant must rotate along the Reeb vector and it can sit at any point in
$y_i$.  Its motion in the phase space $(q^A,p^A)$ is characterized by
six free real parameters that are the initial conditions on the
Sasaki-Einstein space plus $R$. Altogether these parameters
reconstruct a copy of the Calabi-Yau and the induced symplectic form on
the phase space reduces to the natural symplectic form of the Calabi-Yau
cone \cite{MSg}.

In the case of the deformed theory, $\Phi$ is a non trivial function
of $y_i$ and the conditions (\ref{eqsmotion}), (\ref{solother}) 
select a subvariety of the internal space. 
Since $e^{-\Phi} =\sqrt{1+\gamma^2 h}$ we  can write the conditions 
for the vanishing of $\Phi$ and $\partial_{y_i}\Phi$ as
\beq 
h=0 \, , \qquad \partial_{y_i} h =0 \, .
\eeq
Here $h$ is the determinant of the two-torus metric which vanishes exactly
on the edges of the polyhedral cone where the torus degenerates. 
In addition its derivative also vanishes 
on the edges as equation (\ref{edges}) clearly shows. 
We see that the BPS condition restricts the dual giant to live on the
$d$ edges of the cone. 

We still have to find the angular direction of rotation of a BPS dual giant,
which is characterized by the conditions $H=1$, $\partial_{y_i} H=0$.
We still expect our giant to rotate along the Reeb vector. We can
compute the value of $H$ for a giant rotating along the Reeb vector
\beq
H=g(K,K)= G + 9(1-G)(g_{33}- h^{ab}g_{a3}g_{b3})=  \frac{1 + 9\gamma^2 \det{g_{ij}}}{1 + \gamma^2 h}\, .
\eeq
We can easily check that along an edge where $h=\partial_{y_i}h=0$ 
we have $H=1, \partial_{y_i} H=0$ thus solving the remaining equations of
motion and BPS conditions.

Summarizing, a dual giant graviton in the beta-deformed theory is supersymmetric
only when it lives on the edges of polyhedron and rotates along the Reeb vector.
 
Adding $R$ to the set of initial conditions of the probe, we see that 
the moduli space for a dual giant can be identified with  
a collection of lines. We expect that the classical phase space of
a single dual giant corresponds to 
the abelian moduli space of the dual gauge theory. 
Indeed what we found is consistent with the results for static probes and
the field theory discussion in Section \ref{gaugetheory}.
  
\subsubsection{D5 dual giant gravitons}
\label{d5dualg}
For $\gamma$ rational another class of brane probes can be consistently
embedded in the deformed geometry: D5-branes wrapping the same $S^3$ inside
$AdS_5$ and the two-torus spanned by $(\phi^1, \phi^2)$ in the 
internal manifold. The corresponding embedding is 
\bea
\label{d5embedding}
&& t = \tau, \quad R= R(\tau), \quad \theta = \zeta^1, \quad \alpha_1 = \zeta^2,
\quad \alpha_2 = \zeta^3 \, ,\nn\\
&& \phi^1 = \zeta^4, \quad \phi^2 = \zeta^5, \nn\\
&& \phi^3 = \phi^3(\tau), \quad  y_i = y_i(\tau) \qquad i = 1,2,3 \, ,
\eea
where we call $(\zeta^0, \ldots, \zeta^5)$ the 
world-volume coordinates on the brane.
The discussion is completely parallel to that for a static D5-brane.
The world-volume action for the dual giant is still given by (\ref{d5wvaction1})
and now the pulled-back metric is given by
%
\beq
\label{d5pbmetric}
\left( \begin{array}{cccccc}
- \Delta & 0  & 0  & 0  & G \, \dot{\phi}^3 g_{1 \, 3} 
& G \dot{\phi}^3 g_{2 \, 3} \\
0 & R^2 & 0  & 0  & 0 & 0  \\
0 & 0  & R^2 \cos^2 \zeta^1 & 0 & 0 & 0  \\
0 & 0  & 0  & R^2 \sin^2 \zeta^1 & 0 & 0 \\
G \,  \dot{\phi}^3 g_{1 \, 3} & 0  & 0  & 0  & G \, h_{1 \, 1} &  G \, h_{1 \, 2}  \\
G \, \dot{\phi}^3 g_{2 \, 3} & 0 & 0 &0  & G \, h_{2 \, 1} &  G \, h_{2 \, 2}
\end{array} \right)
\eeq
with $ \Delta = 
V(R) - \frac{\dot{R}^2}{V(R)} - g^{i \, j} \dot{y}_i 
\dot{y}_j + \tilde{g}_{3 \, 3} (\dot{\phi}^3)^2 $.
The B-field is given by
\bea
\label{d5pbBfield}
&& B_{04} = - \gamma \, h \, G (h^{2 \, a} g_{a 3}) \dot{\phi}^3 \, ,\\
&& B_{05} =  \gamma \, h \, G (h^{1 \, a} g_{a 3}) \dot{\phi}^3 \, ,\\
&& B_{45} =  \gamma \, h \, G \, ,
\eea
and the world-volume field strength has both magnetic and electric components
\beq
\label{d5fs}
{\rm F}_{45} = \frac{1}{\gamma}  \, ,\qquad {\rm F}_{04}(\tau)  \, , \qquad {\rm F}_{05}(\tau) \, .
\eeq
It is a straightforward computation to verify that
the BI action for the D5 probe has the same form as for the Calabi-Yau
case\footnote{$S_{BI}$ and $S_{WZ}$ are proportional to $T_5 L^4 \alpha'
  Vol(S^3) Vol(T^2)= \pi^4 N / Vol(X_5)$. Again we write only the
  factor $N$.}
\beq
\label{d5BI}
S_{BI} = - \frac{N}{\gamma} \int {\rm d}t R^3 
\sqrt{V(R) - \frac{\dot{R}^2}{V(R)} - g^{i \, j} \dot{y}_i 
\dot{y}_j - g_{3 \, 3} (\dot{\phi}^3)^2 + 2 \gamma g_{3 \, a } \dot{\phi}^3 \hat f_a - \gamma^2 h_{a \, b}  \hat f^a \hat f^b } \, ,
\eeq
where $\hat f^a = \epsilon_{ab} {\rm F}_{0b}$.
The Wess-Zumino part of the action reduces to the Calabi one as well.
This is because the only non trivial contribution is
\beq
\label{d5WZ}
S_{WZ} = T_5 \int C_4 \wedge {\rm }F_{45} = \frac{N}{\gamma} \int {\rm d}t R^4 \, .
\eeq

Thus the world-volume Lagrangian is
\beq
\label{d5lwv}
{\cal L} = - \frac{ N R^3}{\gamma} (\sqrt{\Sigma} - R )
\eeq
with
\beq
\label{sigma}
\Sigma = V(R) - \frac{\dot{R}^2}{V(R)} - g^{i \, j} \dot{y}_i 
\dot{y}_j - g_{3 \, 3} (\dot{\phi}^3)^2  + 2 \gamma g_{3 \, a } \dot{\phi}^3 \hat f_a - \gamma^2 h_{a \, b}  \hat f^a \hat f^b 
\eeq
which formally is equivalent to that of a D3 dual giant in the undeformed geometry 
with the replacement of $\dot\phi^a$ with $-\gamma \epsilon^{ab} {\rm F}_{0b}$. 
On the undeformed Calabi-Yau a D3 dual giant can live at an arbitrary point and 
rotates along the Reeb vector.
We thus see that a class of solutions for D5 dual giants is obtained by 
choosing
\begin{equation}
{\rm F}_{0a}= \frac{1}{\gamma} \epsilon_{ab} b^b \, ,\qquad \dot \phi^3 = b^3 \, .
\label{sol}
\end{equation}

We can analyse the classical phase space of the D5 dual giants.
Exactly as in the case of static D5, for $\beta=m/n$, we obtain the
orbifold ${\rm CY} / \mathbb{Z}_n\times \mathbb{Z}_n$.  Coordinates on this
space are obtained by adding $R$ to the initial values of $\phi^3$,
$y_i$ and the two Wilson lines along the two-torus, and taking into
account the modified periodicities of the angles. The classical phase
space of the D5 dual giants is thus isomorphic to the additional Higgs
branches in the moduli space of the dual gauge theory existing for
rational $\beta$. This is consistent with the fact that the
quantisation of this classical phase space (as done for example in
\cite{MSg}) should reproduce the mesonic BPS operators parameterising
the Higgs branch.

\section{Supersymmetric D-brane probes from \qquad\qquad\,\, 
$\beta$-transformation} 
 
In this section we analyse the existence and supersymmetry of D3 and D5 probes
using generalised geometry. We show in particular 
that the class of dual giants found in Section
\ref{dualgiant} can be obtained by direct action of the
$\beta$-transformation on the word-volume of the D3 dual giants
described in \cite{MSg}. This will automatically ensure that the dual giants are
supersymmetric in the $\beta$-deformed background.

A simple way to do it is again using the formalism of Generalised
Geometry, where a D-brane wrapping a submanifold $\Sigma$ and
supporting a world-volume field strength ${\rm F}$ is described by its
generalised tangent bundle $T_{(\Sigma,{\rm F})}$
\cite{gualtieri}. This can be described as a maximally isotropic
subspace of $T\oplus T^\star$ \footnote{Strictly speaking we should
  consider the extension of $T$ by $T^{\star}$; for our class of
  backgrounds the two are isomorphic since $B$ is globally defined.},
as follows
\bea\label{gentang}
T_{(\Sigma,{\rm F})}=\{X+\xi\in T \oplus T^{\star}|_\Sigma\ :\  X\in T_\Sigma\ {\rm and}\  \xi|_\Sigma=\iota_X{\rm F}    \} \, .
\eea     
As already mentioned, the elements of $T \oplus T^{\star}$ transform
linearly under the action of the extended T-duality group $O(d,d)$ and
so does $T_{(\Sigma,{\rm F})}$. If we start from a D-brane preserving
a background supersymmetry which is also preserved by the $O(d,d)$
transformation, then the D-brane obtained by `integrating' the
transformed generalised tangent bundle will be automatically
supersymmetric in the transformed background.

Let us start by considering the $\beta$-deformation of a
static D3-brane in the undeformed toric Sasaki-Einstein background,
filling the four Poincar\'e directions and sitting at an arbitrary
point of the internal Calabi-Yau cone. As it is well known, this
configuration preserves all the background Poincar\'e supersymmetries.

If the D3-brane sits at a point where the two-torus $(\phi^1,\phi^2)$
shrinks to zero size, the generalised tangent bundle describing the
new D-brane is identical to the one we started from, since the
$\beta$-transformation (\ref{OLM}) reduces to the identity at these
points. Thus the original D3-brane is mapped to a D3-brane at the same
degeneration point in the deformed background.

The situation is different when the original D3-brane sits at a point
where $\phi^a$ are non-degenerate.  Since the only coordinates playing
a non-trivial role in the $\beta$-transformation are the two angles
$\phi^a$ we can simply describe the D3-brane as a point on the
two-torus $(\phi^1,\phi^2)$. Since all forms vanish when restricted to a point,
 the associated (two-dimensional)
generalised tangent bundle (\ref{gentang}) admits the basis $e^a=\d
\phi^a$. Acting on this basis with the $\beta$-deformation
(\ref{OLM}), we obtain a basis for the $\beta$-transformed generalised
tangent bundle
\bea\label{newbase}
\tilde e^a= - \gamma \epsilon^{ab}\frac{\partial}{\partial\phi^b}+\d\phi^a\ .
\eea
By projecting it onto the background tangent bundle, we see that the
ordinary tangent bundle of the new D-brane is spanned by
$\partial_{\phi^1}$ and $\partial_{\phi^2}$. Thus, we obtain a
D5-brane wrapping $(\phi^1,\phi^2)$ in the $\beta$-deformed
background. From the general definition (\ref{gentang}), we also see
that the D5-brane must support a world-volume gauge field ${\rm
  F}=(1/\gamma)\d \phi^1\wedge\d\phi^2$.

We can easily check this result using the supersymmetry conditions for
D-branes given in terms of the (twisted) background pure-spinors
\cite{Martucci,lucasup}.  For a D-brane wrapping the internal cycle
$\Sigma$ with world-volume flux ${\rm F}$ is
\bea
&& [\hat\Psi_-|_\Sigma\wedge e^{\rm F}]_{{\rm top}-1}=0\,\, ,\quad [(\iota_X\hat\Psi_-)|_\Sigma\wedge e^{\rm F}]_{\rm top}=0
\quad \forall X\in T_M \,\, \text{(F-flatness)}\label{F-flat}\\
&& [\hat\Psi_+|_\Sigma\wedge e^{\rm F}]_{{\rm top}}=0\ .\hspace{7.1cm} \text{(D-flatness)}\label{D-flat}
\eea     

In our case $\hat\Psi_-=e^\beta\cdot (e^{-3 A} \Omega^{(0)})$ and
$\hat\Psi_+=e^\beta\cdot\exp(-ie^{-2A}J^{(0)})$. Then, we immediately
see that a D3-brane is supersymmetric only where $\beta\rightarrow 0$
(i.e. the points where the $(\phi^1,\phi^2)$ two-torus degenerates),
since at the other points the F-flatness is not satisfied. On the
other hand, a D5-brane wrapping the $(\phi^1,\phi^2)$ two-torus at any
non-degenerate point automatically satisfies the D-flatness, since
$J^{(0)}|_{T^2}=0$, while the F-flatness imposes the condition
$\text{F}=(1/\gamma)\d\phi^1\wedge\d\phi^2$.
We have thus recovered the result obtained from T-duality,
generalising the result obtained by other means in \cite{LM} for
$AdS_5\times S^5$.

\bigskip

Let us now pass to the description of the action of the
$\beta$-transformation on the D3 dual giant gravitons. D3 dual giants 
in the undeformed background have been found and discussed in
\cite{MSg}.  In any toric Sasaki-Einstein background, they wrap a
static $S^3$ of arbitrary radius at the center of $AdS_5$, sit at any
point described by the $y_i$ coordinates (constrained by the condition
$2b^iy_i=1$) and run along the angular coordinates as follows
\bea\label{MSemb}
t=\tau\quad,\quad \phi^i = b^i \tau + \textrm{const}\ .
\eea  

As for the case above, if a D3 dual giant sits at a point in the $y_i$
coordinates where the two-torus described by $(\phi^1,\phi^2)$
degenerates, its $\beta$-transformation is trivial and gives again a
D3 described by the same embedding (\ref{MSemb}). These are nothing
but the D3-brane dual giants described in Subsection \ref{d3dualg},
which are thus supersymmetric.

In order to study the $\beta$-transformation of D3 dual giants sitting
at non-degeneration points, we can restrict our attention on the time
$t$ and the three angles $\phi^i$. From (\ref{gentang}) we see that a
basis for the generalised tangent bundle of these D3 dual giants is
given by the tangent vectors and a basis of one forms vanishing along the trajectory
\bea\label{MSbase}
e^0=\frac{\partial}{\partial\tau}=\frac{\partial}{\partial t} +
b^i\frac{\partial}{\partial \phi^i}\quad,\quad e^3 = \d t - g_{ij} b^j
\d \phi^i \quad,\quad e^\alpha=c_{(\alpha)i}\d\phi^i\ , 
\eea 
where
$\alpha=1,2$, $i,j=1,2,3$ and $c_{(\alpha)i}$ are such that
$c_{(\alpha)i}b^i=0$. By $\beta$-transforming it 
\bea\label{d5dual}
&&\tilde e^0=\frac{\partial}{\partial t} + b^i\frac{\partial}{\partial
  \phi^i}\quad,\quad \tilde e^3=\gamma \epsilon^{ab} g_{aj} b^j
\frac{\partial}{\partial \phi^b} + \d t - g_{ij} b^j \d\phi^i\ ,\cr &&
\tilde
e^\alpha=\gamma\epsilon^{ab}c_{(\alpha)b}\frac{\partial}{\partial\phi^a}+
c_{(\alpha)i}\d\phi^i\ .  
\eea 
Projecting this basis to the background
tangent bundle we obtain a basis for the tangent bundle to the
$\beta$-transformed brane, which is thus a D5-brane described by the
embedding 
\bea (\tau,\sigma^a)\quad\mapsto\quad (t=\tau\ ,\ \phi^3 =
b^3 \tau + \textrm{const} \ ,\ \phi^a=\sigma^a)\ .  
\eea 
As above,
from the `twisting' of the basis (\ref{d5dual}) we see that the
D5-brane must support a non-trivial world-volume field strength, which
can be easily calculated to be 
\bea\label{fs}
  {\rm F}=\frac{1}{\gamma}\Big(\epsilon_{ab}b^b\d\tau\wedge\d\phi^a+\d\phi^1\wedge\d\phi^2\Big)= 
\frac{1}{2\gamma}\epsilon_{ab}\Big(- b^a \d\tau+\d\phi^a\Big)\wedge
\Big(- b^b \d\tau+\d\phi^b\Big)\ .  
\eea 
We have thus recovered
the D5 dual giants described in Subsection \ref{d5dualg}. Again, they
are automatically supersymmetric by $O(2,2)$ symmetry. As already discussed in Section
\ref{static}, the gauge field must be quantised, giving the condition $\gamma = m/n$ rational.

\bigskip

In Sections \ref{static} and \ref{d5dualg} we showed that the moduli
space of D5-brane probes (static or dual giants) is given by ${\rm CY}/
\mathbb{Z}_n\times \mathbb{Z}_n$. Here we will briefly show that the
same result can be obtained as the $\beta$-deformation of the moduli
space of a probe D3 in the undeformed geometry.

For simplicity, consider a static D3-brane in an undeformed
Sasaki-Einstein background (the analysis of dual giants is completely
analogous).  As explained in \cite{lucasup}, the
infinitesimal deformations of a D-brane wrapping a cycle $\Sigma$ with field
strength ${\rm F}$ are described by sections of the generalised normal
bundle: ${\cal N}_{(\Sigma,{\rm
    F})}=E|_\Sigma/T_{(\Sigma,{\rm F})}\simeq T^\star_{(\Sigma,{\rm
    F})}$. In the case of the static D3-brane, focusing again on the
$(\phi^1,\phi^2)$ directions, a basis for the sections of ${\cal
  N}_{(\Sigma,{\rm F})}$ is given by the following representatives
\bea
e_a=\frac{\partial}{\partial \phi^a}\ ,
\eea
which clearly generate the motion of the D3-brane in the
$(\phi^1,\phi^2)$ directions. We can now apply the
$\beta$-transformation (\ref{OLM}) to obtain representatives of the
corresponding sections of the generalised normal bundle to the
D5-brane in the $\beta$-deformed background. The are given by
 \bea
 \tilde e_a=\frac{1}{\gamma}\epsilon_{ba}\d\phi^b\ .
 \eea  
The displacement 
\bea
\phi^a\rightarrow \phi^a+c^a 
\eea
of the D3-brane in the Sasaki-Einstein background is generated by the generalized normal vector $c^ae_a$. The $\beta$-transformation maps it into $c^a\tilde e_a$, which corresponds, as discussed in \cite{lucasup}, to a shift $\Delta A=c^a\tilde e_a$ of the gauge field on the D5-brane in the $\beta$-deformed background. In components this reads
\bea
A_a\rightarrow A_a + \frac{1}{\gamma}\epsilon_{a b}c^b=A_a + n\epsilon_{a b}c^b
\eea

Thus, in particular, a periodic shift $\Delta_a\phi^b=2\pi\delta^b_a$ of the D3-brane corresponds to a shift
\bea\label{wshift}
\Delta_a\int_b \mathcal{A} = 2 \pi n \epsilon_{ba}\ 
\eea
of the Wilson line on the D5-brane. As before the Wilson lines are defined by  $\int_a \mathcal{A}$, with
$\mathcal{A}= A/ 2 \pi$, have period $2 \pi$ and parameterise a two-torus $\tilde T^2$.

 This result have a natural
interpretation taking into account that the $\beta$-deformation
maps $n$ D3-branes to a single D5-brane. From this point of view, the
angular positions $\phi^a$ in the undeformed background actually corresponds to the average
$\langle\phi^a\rangle=\sum_{r=1}^n \phi^a_{(r)}/n$ of the angular
positions $\phi^a_{(r)},r=1,\ldots,n,$ of the $n$ D3-branes, while the
Wilson lines on the D5-brane in the $\beta$-deformed background are
associated to the sums $\sum_{r=1}^n \phi^a_{(r)}$ (the trace of the
corresponding $n\times n$ matrix in the complete non-abelian
description of the $n$ D3-branes) by the $\beta$-deformation. A
constant periodic shift $\Delta_a\langle\phi^b\rangle=2\pi\delta^b_a$
of the average D3-brane position then produces the shift
(\ref{wshift}) of the D5-brane Wilson lines.
From (\ref{wshift}), we see that going once around a 1-cycle in
$T^2_{\rm SE}$ corresponds to going $n$-times around a 1-cycle in
$\tilde T^2$
\bea
\tilde T^2\simeq T^2_{\rm SE}/(\mathbb{Z}_n\times \mathbb{Z}_n)\ .
\eea
We can conclude that the moduli space of the static
D5-branes in the $\beta$-deformed background corresponds to the quotient CY$/(\mathbb{Z}_n\times
\mathbb{Z}_n)$ of the CY cone of the undeformed theory. The same
arguments presented above can be applied to the case of D5 dual giants
in the $\beta$-deformed background and lead to the expected conclusion
that their moduli space again corresponds to CY$/(\mathbb{Z}_n\times
\mathbb{Z}_n)$.

However, until now we have given only a one-to-one map between the coordinates on the moduli space
and the coordinates on CY$/(\mathbb{Z}_n\times \mathbb{Z}_n)$. To
complete the identification we still have to compute the metric on the
moduli space and see that it coincides with the metric of
CY$/(\mathbb{Z}_n\times \mathbb{Z}_n)$.

Consider the moduli space of a static supersymmetric D5-brane
described above. Its tangent vectors correspond to the fluctuations
in the internal space that preserve the supersymmetry condition and
can thus be seen as massless chiral fields in an effective
four-dimensional description. The K\"ahler metric for these chiral
fields can be in principle obtained by looking at their kinetic term
obtained by expanding the DBI+CS action for the D5-brane. This is
exactly the metric we are interested in.

We can apply the results of
\cite{lucasup,lucapaul1} to identify the K\"ahler
structure of the moduli space. To find the correct
holomorphic parametrization of the D5 massless fluctuations we can use once again the
action of the $\beta$-deformation. 
The fluctuation of a general
D-brane are given by the sections of the generalised normal bundle
$\caln_{(\Sigma,{\rm F})}$ \cite{lucasup}. For a D3-brane in a
Sasaki-Einstein background, the moduli space corresponds to the CY
cone $M$ itself, $\caln_{(\Sigma,{\rm F})}\equiv T_M$ and the
associated complex structure is nothing but the complex structure of
the CY. Now, a basis for the holomorphic tangent space to the moduli
space is given by the following sections of the generalised normal
bundle
\bea\label{holbased3}
e_i=\frac{\partial}{\partial z^i}\ ,
\eea 
where $z^i$ are the holomorphic coordinates on the CY.  A basis for
the holomorphic deformations for the corresponding D5-brane in the
$\beta$-deformed background can be obtained simply by taking the
$\beta$-transformation of the basis (\ref{holbased3})
\bea\label{holbased5}
\tilde e_i=O_{LM}\cdot e_i\ .
\eea  

We can now use the general formula for the K\"ahler metric given in
\cite{lucasup,lucapaul1}, which was in fact obtained by expanding the
DBI+CS D-brane action. In the basis (\ref{holbased5}) it is given by
\bea
\calg_{i\bar\jmath}&=&-i\int_\Sigma[\tilde e_i\cdot \bar{\tilde e}_{\bar\jmath}\cdot 
\Im (e^{2A}\hat\Psi_+)]|_\Sigma\wedge e^{\text{F}}=\cr
&=& -i\int_\Sigma\big\{e^{2A}e^\beta\cdot\iota_{e_i}\iota_{\bar{e}_{\bar\jmath}}
\Im\big[ \exp(-ie^{-2A}J_{(0)})\big]\big\}|_\Sigma\wedge e^{\text{F}}=\cr
&=& -iJ^{(0)}_{i\bar\jmath}\int_\Sigma \text{F}= -i (2\pi)^2 n J^{(0)}_{i\bar\jmath}\ ,
\eea
where $J^{(0)}$ is K\"ahler form on the CY cone.  We thus see that we
obtain (locally) exactly the CY metric, up to an overall factor which
comes from the fact that the D5-brane with $n$ units of $\text{F}$
flux corresponds to $n$ D3-branes in the undeformed SE
background. From the coordinate identification discussed above, we can
conclude that the K\"ahler moduli space for the D5-brane is indeed
CY$/(\mathbb{Z}_n\times \mathbb{Z}_n)$.

\section{Comments on giant gravitons}

There exist other BPS string configurations. Of particular
interest are the giant gravitons, configurations of D3-brane
wrapping 3 cycles in the internal space. It would be quite
interesting to perform a complete analysis of the spectrum of giant
gravitons on the $\beta$-deformed background.  As shown in
\cite{Beasley,Minwalla,FHZ,Forcella:2007wk,BVFHZ,Forcella:2007ps}, in the undeformed case, the
quantisation of the classical supersymmetric giant graviton solutions
gives a complete information about the spectrum and the partition
function of BPS mesonic operators in the field theory.

In the Calabi-Yau case, giant gravitons can be parameterised by
Euclidean D3-branes living inside the internal six-manifold
\cite{Mikhailov,Beasley}.  We restrict to the minimal giant gravitons
without world-volume flux, which parametrize all the bosonic BPS
states. The argument given in \cite{Beasley} suggests that the same
parameterisation can be used in all solutions with $AdS_5$ factor.
The supersymmetric conditions for Euclidean D-branes on a generalised
geometry background have been derived in \cite{lucapaul2} and shown to
be identical to the conditions for the internal part of space-filling
branes discussed in \cite{Martucci,lucasup}\footnote{Indeed, the results of this section can be equally used to identify and study flavor D7-branes on this general class of $\beta$-deformed backgrounds (see \cite{Mariotti:2007ym,penati} for work in this direction).}, that we have already
written in (\ref{F-flat}) and (\ref{D-flat}).  So they can be easily
applied to an Euclidean D3-brane, given the form of the pure spinors
discussed in Section \ref{pure}.


The F-flatness condition (\ref{F-flat}) for Euclidean D3-brane wrapping $\Sigma$ with $\text{F}=0$ reduces to
\bea\label{hold3}
\Omega_{(0)}|_\Sigma=0\ ,
\eea
where we recall that $\Omega_{(0)}$ is the holomorphic $(3,0)$ on the
original CY geometry.  The condition (\ref{hold3}) exactly requires
that the 4-cycle wrapped by the Euclidean D3-brane must be holomorphic
with respect to the CY complex structure. Consider for example
four-cycles in $\beta$-deformed toric vacua defined by the embedding
$w_3=g(z^1,z^2,\bar z^1,\bar z^2)$, where $z^{1,2},\bar z^{1,2}$ are
chosen as coordinates on the cycle. Then the F-flatness (\ref{hold3})
becomes
\bea
\d z^1\wedge \d z^2\wedge \d g=0 \quad\Leftrightarrow\quad \bar\partial g=0\ ,
\eea
which indeed requires that the embedding is holomorphic with respect
to the old variables. Of course, other supersymmetric embeddings might
exist which are not parameterised by $z^{1,2}$.


On the other hand, the general D-flatness condition is (\ref{D-flat})
in the $\beta$-deformed toric-vacua, for the above four-cycles with
$\rm{F}=0$, becomes
\bea
\iota_\beta(J\wedge J\wedge J)|_\Sigma&\sim& \d x^1\wedge \d x^2\wedge \d g\wedge \d\bar g=0\quad\Leftrightarrow\quad \Im (\partial_{1}g\,\bar\partial_{\bar 2}\bar g)=0\ .
\eea


Interestingly, all the supersymmetric conditions can be written in
terms of the original complex coordinates of the Calabi-Yau. This is
in agreement with field theory, where the moduli space for the
deformed theory remains a complex manifold and the original complex
structure of the moduli space can be still used to characterize it.
We can easily find many solutions of the F and D-flatness
conditions. For example, all {\it monomials} of definite charge 
 $w_3=e^{n_1 z_1}e^{n_2 z_2}$ solve
the constraints.  At first sight, we are left
with more solutions than expected from the spectrum of BPS states of
the deformed theory. However a more careful analysis of the giant
graviton characterization as Euclidean D3-branes, of their global
properties, of their world-volume flux and, in general,  of 
the quantisation procedure should be performed before
extracting correct results. We leave this interesting analysis for
future work.

\section{The gauge theory}\label{gaugetheory}
In this Section we discuss the moduli space for a $\beta$-deformed quiver gauge 
theory. Rather than giving general proofs for all toric quiver theories we 
examine various examples and we give some general arguments.

\subsection{Non abelian BPS conditions}\label{nonab}
In order to understand the full mesonic
 moduli space of the gauge theory we need 
to study general non-abelian solutions of the F term equations. 

Before attacking the general construction, 
we consider ${\cal N}=4$ SYM and the conifold. 
In the ${\cal N}=4$ SYM case, we form mesons out of the three adjoint fields $(\Phi_i)_\alpha^\beta$. The non-abelian BPS conditions for these mesonic fields are given in equation (\ref{N4}) and can be considered as equations
for three $N\times N$ matrices. In the conifold case, we can 
define four composite mesonic fields which transform in the adjoint 
representation of one of the two gauge groups
\begin{equation}
x=(A_1B_1)_\alpha^\beta,\,\,\,\, y=(A_2B_2)_\alpha^\beta,\,\,\,\, z=(A_1B_2)_\alpha^\beta,\,\,\,\,  w=(A_2B_1)_\alpha^\beta
\end{equation}
and consider the four mesons $x,y,z,w$ as $N\times N$ matrices. We could use 
the second gauge group without changing the results. With a simple 
computation using the F-term conditions (\ref{conFterm}) we derive the
following matrix commutation equations
\begin{eqnarray}
\label{comm}
x z &=& b^{-1} z x\nonumber\\
x w &=& b w x\nonumber\\
y z &=& b z y\nonumber\\
y w &=& b^{-1} w y\nonumber\\
x y &=& y x \nonumber\\
z w &=& w z
\end{eqnarray}
and the matrix equation
\begin{equation}
\label{eq} 
x y = b w z 
\end{equation}
which is just the conifold equation. Here and in the following
 $b=e^{-2 i \pi \beta}$. For $\beta=0$ these conditions simplify.
All the mesons commute and the $N\times N$ matrices $x,y,z,w$ can be simultaneous diagonalized. The eigenvalues are required to satisfy the conifold equation
(\ref{eq}) and therefore the moduli space is given by the symmetrized
product of $N$ copies of the conifold, as expected. 

An interesting observation is that, for the ${\cal N}=4$ SYM 
and (\ref{comm}) for the conifold, the F-term conditions for 
$\beta \ne 0$ can be obtained by using the non commutative product
defined in (\ref{star})
\begin{equation}\label{star2}
f*g \equiv e^{i \pi \beta (Q^f \wedge Q^g)} f g  \, .
\end{equation} 
The charges of mesons for ${\cal N}=4$ and the
conifold are shown in Figure \ref{n4con}.

The BPS conditions for the Calabi-Yau case, which require that every 
pair of mesonic fields
$f$ and $g$ commute, are replaced in the $\beta$-deformed theory by a
non commutative version 
\beq \label{starcomm} [f,g]\, =\, 0  \qquad \rightarrow \qquad [f,g]_\beta\equiv f*g-g*f \, =\, 0 \, . \eeq
It is an easy exercise, using the assignment of charges shown in Figure \ref{n4con}, to show that these modified commutation relations reproduce
equations (\ref{N4}) and (\ref{comm}).


This simple structure extends to a generic toric gauge theory. The algebraic equations of the Calabi-Yau give a set of matrix equations for mesons. In the undeformed theory, all mesons commute, while in the $\beta$-deformed theory the
original commutation properties are replaced by their non commutative version. 
In order to fully appreciate these statements we need to understand the structure
of the mesonic chiral ring for toric theories \cite{vegh,MSY2,butti,BFZ,benve,Hanany:2006uc,Feng:2007ur}.

\subsubsection{The mesonic chiral ring}
We briefly review the structure of the mesonic chiral ring for quiver gauge 
theories. The reader is referred to \cite{vegh,MSY2,butti,BFZ,benve,Hanany:2006uc,Feng:2007ur} for an exhaustive discussion. The reader who wants to avoid
technical details can directly jump to the next Sections, where most of
the examples are self-explaining.

\begin{figure}[ht]
\begin{center}
\includegraphics[scale=0.70]{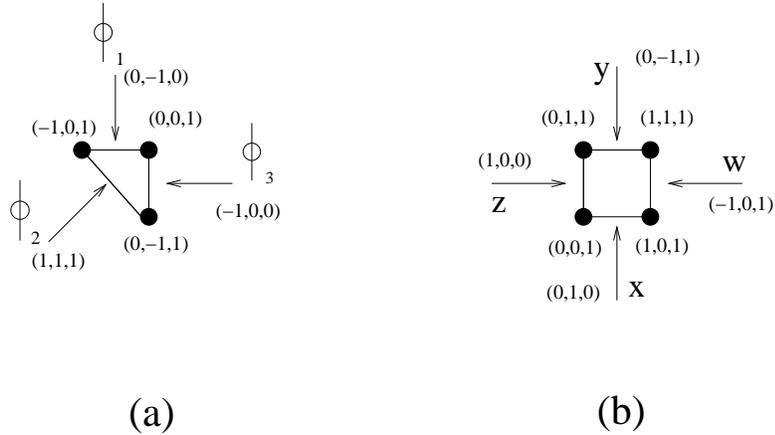} 
\caption{The toric diagram $\mathcal{C}$ and the generators of the dual cone $\mathcal{C}^*$ with the associated mesonic fields for: (a) $\mathcal{N}=4$, (b) conifold. The $U(1)^3$ charges of the mesons are explicitly indicated; the first two
entries of the charge vectors give the $U(1)^2$ global charge used to define
the non commutative product.}
\label{n4con}
\end{center}
\end{figure}

>From the algebraic-geometric 
point of view  the data of a conical toric Calabi-Yau are encoded in a 
rational polyedral cone  $\mathcal{C}$ in $\mathbb{Z}^3$ defined by a set of
vectors $V_{\alpha}$ $\alpha=1,...,d$. 
For a CY cone, using an $SL(3, \mathbb{Z})$ transformation, 
it is always possible to carry these vectors in the form $V_{\alpha}=(x_{\alpha},y_{\alpha},1)$. 
In this way the toric diagram can be drawn in the $x,y$ plane (see for example 
Figure \ref{n4con}). 
The CY equations can be reconstructed from this set of combinatorial data using the dual cone $\mathcal{C}^*$. 
This is defined in equation (\ref{cone}) 
and it was already used to write the metric as a $T^3$ fibration.  
The two cones are related as follow. The geometric generators for 
the cone $\mathcal{C}^*$, which are vectors aligned along the edges of $\mathcal{C}^*$, are the perpendicular vectors to the facets of 
$\mathcal{C}$.

To give an algebraic-geometric description of the CY, we 
need to consider the cone $\mathcal{C}^*$ as a semi-group and to find 
its generators over the integer numbers. The primitive vectors pointing along 
the edges generate the cone over the real numbers but we generically need 
to add other vectors to obtain a basis over the integers. Denote by 
$W_j$ with $j=1,...,k$ a set of generators of $\mathcal{C}^*$ over the 
integers.
To every vector $W_j$ it is possible to associate a coordinate $x_j$ in some 
ambient space. $k$ vectors in $\mathbb{Z}^3$ are clearly linearly dependent 
for $k > 3$, and the additive relations satisfied by the generators $W_j$ 
translate into a set of multiplicative relations among the coordinates $x_j$. 
These are the algebraic equations defining the six-dimensional CY cone.

All the relations between points in the dual cone become relations among mesons
in the field theory. In fact,
using toric geometry and dimer technology, it is possible to show that there 
exists a one to one correspondence between the integer points inside
$\mathcal{C}^*$ and the mesonic operators in the dual field theory, modulo 
F-term constraints \cite{MSY2,benve}. 
To every integer point $m_j$ in  $\mathcal{C}^*$
we indeed associate a meson $M_{m_j}$ in the gauge theory with $U(1)^3$ charge
$m_j$.  In particular, the mesons are uniquely determined by their charge
under $U(1)^3$. The first two coordinates 
\beq Q^{m_j}=(m_j^1,m_j^2)\eeq 
of the vector $m_j$ 
are the charges of the meson under the two flavour $U(1)$ symmetries. 
Since the cone $\mathcal{C}^*$ is generated as a semi-group by the vectors $W_j$
the generic meson will be obtained as a product of basic mesons $M_{W_j}$,
and we can restrict to these generators for all our purposes.
The multiplicative relations satisfied by the coordinates $x_j$ become
a set of multiplicative 
relations among the mesonic operators  $M_{W_j}$ inside the chiral ring of the gauge theory. It is possible to prove that these relations are a consequence
of the F-term constraints of the gauge theory.
The abelian version of this set of relations is just the
set of algebraic equations defining the CY variety as embedded in 
$\mathbb{C}^k$.  The examples of ${\cal N}=4$ SYM and the conifold are shown in Figure \ref{n4con}. In the case of ${\cal N}=4$ , the three mesons $\Phi_j$
correspond to independent charge vectors and we obtain the variety 
$\mathbb{C}^3$. In the case of the conifold, the four mesons $x,y,z,w$ 
correspond to four vectors with one linear relation and we obtain the 
description of the conifold as a quadric $xy=zw$ in $\mathbb{C}^4$.

We need now to understand the non abelian structure of the BPS conditions.
Mesons correspond to closed loops in the quiver 
and, as shown in \cite{vegh,butti}, for any meson
there is an F-term equivalent meson that passes for a given gauge group.
We can therefore assume that all meson loops have a base point at a 
specific gauge group and consider them as $N\times N$ matrices 
${\cal M}_\alpha^\beta$. In the undeformed theory,
the F-term equations imply that all mesons commute and can be 
simultaneously diagonalized. 
The additional F-term constraints require that the mesons, and therefore
all their eigenvalues, satisfy the algebraic equations defining the Calabi-Yau.
This gives a moduli space which is the $N$-fold symmetrized
product of the Calabi-Yau. This has been explicitly verified in \cite{grant} 
for the case of the quiver theories \cite{YLgauge} corresponding to the $L^{pqr}$ manifolds. 
In the $\beta$-deformed theory
the commutation relations among mesons are replaced by $\beta$-deformed commutators
\beq\label{mesb}
M_{m_1} M_{m_2}=e^{-2 i \pi \beta (Q^{m_1} \wedge Q^{m_2})} M_{m_2} M_{m_1}=b^{(Q^{m_1} \wedge Q^{m_2})} M_{m_2} M_{m_1} \, .
\eeq
The prescription (\ref{mesb}) will be our short-cut for computing the relevant quantities we will be interested in. 
This fact becomes computationally relevant in the generic toric case.
As we will show in an explicit example in the Appendix \ref{apx} this procedure is equivalent to using 
the $\beta$-deformed superpotential defined in (\ref{defSup}) 
and 
deriving the constraints for the mesonic fields from the F-term relations.

Finally the mesons still satisfy a certain number of 
algebraic equations
\begin{equation}\label{modCYb}
 f({\cal M})=0 
\end{equation}
which are isomorphic to the defining equations of the original Calabi-Yau. 



\subsection{Abelian moduli space}

In this section, we give evidence from the gauge theory 
side that the abelian moduli space of the $\beta$-deformed theories is a set
of lines. There are exactly $d$ such lines, where $d$ is the number of vertices
in the toric diagram.
In fact, the lines correspond to the geometric 
generators of the dual cone of the undeformed geometry, or, in other words, the edges of the polyedron ${\cal C}^*$ where the $T^3$ fibration degenerates
to $T^1$. Internal generators of ${\cal C}^*$ as a semi-group do not correspond
to additional lines in the moduli space. 
These statements are the field theory counterpart 
of the fact that the D3 probes can move only along the edges 
of the symplectic cone.

We explained in the previous section how to obtain a set of modified 
commutation relations among mesonic fields. 
In the abelian case the mesons reduce to commuting c-numbers. 
>From the relations (\ref{mesb}) with non a trivial $b$ factor, we obtain the constraint
\begin{equation}
M_{m_1} M_{m_2}=0 \, .
\end{equation}
Adding the algebraic constraints (\ref{modCYb}) defining the CY, we obtain the full
set of constraints for the abelian mesonic moduli space. 

We now solve the constraints in a selected set of examples, which are 
general enough to exemplify the result. We analyse ${\cal N}=4$, the conifold,
the Suspended Pinch Point ($SPP$) singularity and a more sophisticated
example, $PdP_4$, which covers the case where the generators of ${\cal C}^*$ 
as a semi-group are more than the geometric generators.   

\subsubsection{The case of $\mathbb{C}^3$}
The $\mathcal{N}=4$ theory is simple and was already discussed in 
Section \ref{betadef}. The three lines correspond to the geometric generators
of the dual cone as in Figure \ref{n4con}.

\subsubsection{The conifold}
The abelian mesonic moduli space of the conifold theory was 
already discussed in Section \ref{betadef} using elementary fields. From the 
equations (\ref{comm}) we obtain the same result: four lines corresponding
to the external generators of the dual cone as shown in Figure \ref{n4con}. 

\subsubsection{SPP}
\begin{figure}[ht]
\begin{center}
\includegraphics[scale=0.55]{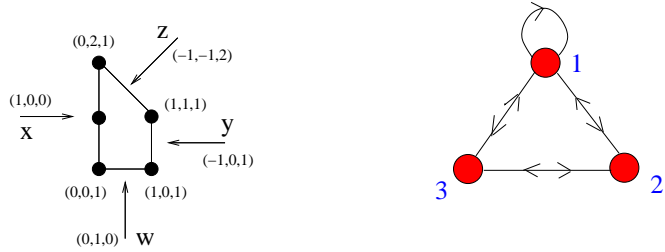} 
\caption{The toric diagram and the quiver of the $SPP$ singularity}
\label{tqspp}
\end{center}
\end{figure}

The gauge theory obtained as the near horizon limit of a stack of D3-branes 
at the tip of the conical singularity
\begin{equation}\label{spp}
x y^2 = w z
\end{equation} 
is called the $SPP$ gauge theory \cite{mp}. The toric diagram and the quiver of this theory are given in Figure \ref{tqspp}. 
Its superpotential is
\begin{equation}\label{superpotspp}
W=X_{21}X_{12}X_{23}X_{32} + X_{13}X_{31}X_{11} -  X_{32}X_{23}X_{31}X_{13} - X_{12}X_{21}X_{11}
\end{equation}
The generators of the mesonic chiral ring are 
\begin{eqnarray}
 & w = X_{13} X_{32} X_{21} \, , & x = X_{11} \, , \nonumber\\
 & z = X_{12} X_{23} X_{31} \, , & y = X_{12} X_{21} \, .
\end{eqnarray} 
These mesons correspond to the generators of the dual cone in Figure 
\ref{tqspp}. Their flavour charges can be read from the dual toric diagram
\begin{equation}\label{messppc}
Q_x=(1,0) \hbox{ , } Q_z=(-1,-1) \hbox{ , }  Q_y=(-1,0) \hbox{ , } Q_w=(0,1) \, .
\end{equation}
Using the deformed commutation  rule for mesons (\ref{mesb}) we obtain the following relations
\begin{eqnarray}\label{sppbc}
& & x w = b w x  \, , \quad z x = b x z  \, , \quad  w z = b z w \, ,\nonumber\\
& & w y = b y w  \, , \quad y z = b z y  \, . 
\end{eqnarray}
In the abelian case they reduce to
\begin{eqnarray}
& & x w = 0 \, , \quad  z x = 0  \, , \quad  w z = 0 \, ,\nonumber\\
& & w y = 0  \, , \quad  y z = 0  \, , \quad  x y^2 \sim w z \, , 
\end{eqnarray}
where the last equation is the additional F-term constraint giving the
original CY manifold.  The presence of the symbol ``$\sim$'' is due to
the fact that the original CY equation is deformed by an unimportant
power of the deformation parameter $b$, which can always be reabsorbed
by rescaling the variables.  The solutions to these equations are
\begin{eqnarray}
( x = 0 \, , \quad y = 0 \, , \quad z=0 ) \rightarrow \{ w \} \, , \nonumber\\
( x = 0 \, , \quad y = 0 \, , \quad w=0 ) \rightarrow \{ z \} \, ,\nonumber\\
( x = 0 \, , \quad z = 0 \, , \quad w=0 ) \rightarrow \{ y \} \, ,\nonumber\\
( w = 0 \, , \quad y = 0 \, , \quad z=0 ) \rightarrow \{ x \} \, , 
\end{eqnarray} 
corresponding to the four complex lines associated to the four generators of the dual cone.

\subsubsection{PdP$_4$}
This is probably the simplest example with internal generators: the
perpendicular to the toric diagram are enough to generate the dual
cone on the real numbers but other internal vectors are needed to
generate the cone on the integer numbers. The discussion in Section
\ref{dualgiant} suggests that the moduli space seen by the dual giant
gravitons and hence the abelian mesonic moduli space of the gauge
theory are exhausted by the external generators. We will see evidence
of this fact.

\begin{figure}[ht]
\begin{center}
\includegraphics[scale=0.55]{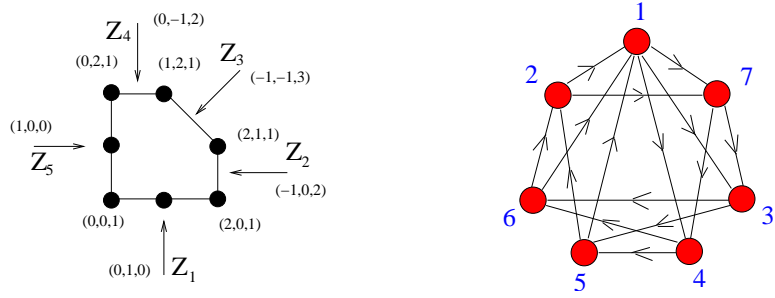} 
\caption{The toric diagram and the quiver of the $PdP_4$ singularity}
\label{PdP4}
\end{center}
\end{figure}

The $PdP_4$ gauge theory, \cite{PdP4}, is the theory obtained as the
near horizon limit of a stack of D3-branes at the tip of the non
complete intersection singularity defined by the set of equations
\begin{eqnarray}\label{P4F}
& & z_1 z_3 = z_2 t \hbox{ ,  } z_2 z_4 =  z_3 t \hbox{ ,  } z_3 z_5  = z_4 t \nonumber\\
& & z_2 z_5 = t^2 \hbox{ ,  } z_1 z_4 =  t^2 \, .
\end{eqnarray} 
The toric diagram and the quiver of the theory are given in Figure \ref{PdP4}.
The superpotential of the theory is
\begin{eqnarray}
W &=& X_{61} X_{17} X_{74} X_{46} + X_{21} X_{13} X_{35} X_{52} + X_{27} X_{73} X_{36} X_{62} + X_{14} X_{45} X_{51} \nonumber\\
&- & X_{51}X_{17} X_{73} X_{35} - X_{21} X_{14} X_{46} X_{62} - X_{27} X_{74} X_{45} X_{52} - X_{13} X_{36} X_{61} \, . 
\end{eqnarray}
The generators of the mesonic chiral ring are
\begin{eqnarray}
& & z_1=X_{51}X_{13}X_{35} \, ,\quad  z_2=X_{51}X_{17}X_{74}X_{45} \, , 
\quad z_3=X_{21}X_{17}X_{74}X_{45}X_{52} \, ,\nonumber\\
& & z_4=X_{14}X_{45}X_{52}X_{21} \, , \quad  z_5= X_{14} X_{46} X_{61} 
\, , \quad  t=X_{13} X_{36} X_{61} \, .
\end{eqnarray}
>From the toric diagram we can easily read the charges of the mesonic generators
\begin{equation}
Q_{z_1}=(0,1) \, ,\quad  Q_{z_2}=(-1,0) \, , \quad   Q_{z_3}=(-1,-1) 
\, , \quad  Q_{z_4}=(0,-1) \, , \quad  Q_{z_5}=(1,0) \, .
\end{equation}
To generate the cone on the integers we need to add the internal generator $t=(0,0,1)$ with flavour charges $Q_t=(0,0)$. 
The generators satisfy the equations (\ref{P4F}) for the $PdP_4$ singularity modified just by some irrelevant proportional factors given by powers of $b$.
We must add the relations obtained from the mesonic $\beta$-deformed 
commutation rule (\ref{mesb})
\begin{eqnarray}
& &z_1 z_2 = b z_2 z_1  \, ,\quad  z_1 z_3 = b z_3 z_1  \, ,\quad
 z_5 z_1 = b z_1 z_5  \, ,\quad  z_2 z_3 = b z_3 z_2 \nonumber\\ 
& &  z_2 z_4 = b z_4 z_2  \, ,\quad  z_3 z_4 = b z_4 z_3  \, ,
\quad  z_3 z_5 = b z_5 z_3  \, ,\quad  z_4 z_5 = b z_5 z_4 \, , 
\end{eqnarray} 
that in the abelian case reduce to 
\begin{eqnarray}\label{P4cb}
& & z_1 z_2 = 0  \, ,\quad z_1 z_3 = 0  \, ,\quad z_5 z_1 = 0  \, ,\quad
 z_2 z_3 = 0 \, ,\nonumber\\
& & z_2 z_4 = 0  \, ,\quad  z_3 z_4 = 0  \, ,\quad  z_3 z_5 = 0  \, ,\quad
 z_4 z_5 = 0 \, . 
\end{eqnarray} 
The solutions to the set of equations (\ref{P4F}) and (\ref{P4cb}) are
\begin{eqnarray}
&& ( z_2 = 0  \, ,\quad  z_3 = 0  \, , \quad  z_4=0  \, ,\quad  z_5 = 0  \, ,\quad   t=0 ) 
 \rightarrow \{ z_1 \} \, , \nonumber\\
&& ( z_1 = 0  \, ,\quad  z_3 = 0  \, , \quad  z_4=0  \, ,\quad  z_5 = 0  \, ,\quad  t=0 ) 
\rightarrow \{ z_2 \} \, ,\nonumber\\
&& ( z_1 = 0 \, ,\quad  z_2 = 0  \, ,\quad  z_4=0  \, ,\quad z_5 = 0  \, ,\quad  t=0 ) 
\rightarrow \{ z_3 \} \, , \nonumber\\
&& ( z_1 = 0 \, ,\quad  z_2 = 0  \, ,\quad z_3=0  \, ,\quad z_5 = 0  \, ,\quad  t=0 ) 
\rightarrow \{ z_4 \} \, , \nonumber\\
&& ( z_1 = 0  \, ,\quad z_2 = 0  \, ,\quad  z_3=0  \, ,\quad  z_4 = 0 \, ,\quad t=0 ) 
\rightarrow \{ z_5 \} \, , 
\end{eqnarray} 
corresponding to the five external generators. We observe in particular that the complex line corresponding to the internal generators $t$ is not a solution.

\subsection{Non abelian moduli space and rational $\beta$}

The F-term equations
\begin{equation}\label{mb}
M_{m_1} M_{m_2}=e^{- 2\pi i \beta (Q^{m_1} \wedge Q^{m_2})} M_{m_2} M_{m_1}
\end{equation}
give a non commutative 't Hooft-Weyl algebra for the 
$N\times N$ matrices ${\cal M}_I$. By diagonalizing the matrix
$\theta_{m_1m_2}=(Q^{m_1} \wedge Q^{m_2})$ we can reduce the problem
to various copies of the algebra for a non commutative torus 
\beq  
M_1 M_2 = e^{2\pi i\theta} M_2 M_1 
\label{eqtheta}
\eeq
 whose representations are well known.

For generic $\beta$, corresponding to irrational values of $\theta$, the 't Hooft-Weyl algebra has no non trivial 
finite dimensional representations: we can only find solutions where
all the matrices are diagonal, and in particular equation (\ref{eqtheta}) implies $M_1 M_2= M_2 M_1=0$. The problem is thus reduced to the
abelian one and the moduli space is obtained by symmetrizing 
$N$ copies of the abelian moduli space, which consists of $d$ lines. 
This is the remaining of the original Coulomb branch
of the undeformed theory. 

For rational $\beta= m/n$, instead, new branches are opening up in the
moduli space \cite{bl,dorey}. In fact, for rational $\beta$,
we can have finite dimensional representations of the 't Hooft-Weyl 
algebra  which are given by $n\times n$ matrices $(O^I)_{ij}$. 
The explicit form of the matrices $(O^I)_{ij}$ can be found in \cite{weyl}
but it is not of particular relevance for us. 
For gauge groups $SU(N)$
with $N=nM$ we can have vacua where the mesons have the form 
\begin{equation}\label{gen}
({\cal M}_I)_\alpha^\beta= {\rm Diag}({\cal M}_a) \otimes (O^I)_{ij},\quad  a=1,...,M,\quad i,j=1,...,n, \quad \alpha,\beta=1\ldots N \, .
\end{equation}
The $M$ variables ${\cal M}_a$ are further constrained by the algebraic 
equations (\ref{modCYb}) and are due to identifications
 by the action of the gauge
group. A convenient way of parameterising the moduli space is to look at
the algebraic constraints satisfied by the elements of the centre 
of the non-commutative algebra \cite{bl}. 

We will give arguments showing that the centre 
of the algebra of mesonic operators is the algebraic variety CY$/\mathbb{Z}_n \times \mathbb{Z}_n$. Here 
CY means the 
original undeformed variety, and the two $\mathbb{Z}_n$ factors are 
abelian discrete sub-groups of the two flavours symmetries.
This statement is the field theory counterpart of the fact that the moduli space of D5 dual  giant gravitons is the original Calabi-Yau
divided by $\mathbb{Z}_n\times \mathbb{Z}_n$. 

The generic vacuum (\ref{gen}) corresponds to $M$ D5 dual giants moving on
the geometry. The resulting branch of the moduli space is the $M$-fold 
symmetrized product of the original Calabi-Yau divided by 
$\mathbb{Z}_n\times \mathbb{Z}_n$. Each D5 dual giant should be considered
as a fully non-abelian solution of the dual gauge theory carrying $n$ color
indices so that the total number of colors is $N=nM$. We can obtain a
different perspective on this branch of our gauge theory by considering
it as the world-volume theory of D3-branes sitting at a discrete torsion
$\mathbb{Z}_n\times \mathbb{Z}_n$ orbifold of the original singularity \cite{discrete}. In this picture,
the D5 dual giants correspond to the physical branes surviving the orbifold
projection. This perspective 
has been discussed in details in the literature for ${\cal N}=4$ SYM \cite{bl} and it can be easily extended to generic toric singularities.

\subsubsection{The case of $\mathbb{C}^3$}
The case of the $\beta$-deformation of $\mathcal{N}=4$ gauge theory is simple and well known \cite{bl}.
 
The generators of the algebra of mesonic operators are the three elementary fields $\Phi_1$, $\Phi_2$, $\Phi_3$. 
Equation (\ref{N4}) implies that it possible to write the generic element of the algebra in the ordered form
\begin{equation}\label{opC3}
\Phi_{k_1,k_2,k_3}=\Phi_1^{k_1} \Phi_2^{k_2} \Phi_3^{k_3}
\end{equation}
The centre of the algebra is given by the subset of operators in (\ref{opC3}) 
such that:
\begin{eqnarray}
\Phi_{k_1,k_2,k_3}\hbox{ }\Phi_1 = b^{k_3 - k_2} \hbox{ }\Phi_1 \hbox{ }\Phi_{k_1,k_2,k_3} = \Phi_1 \hbox{ }\Phi_{k_1,k_2,k_3} \, , \nonumber\\
\Phi_{k_1,k_2,k_3}\hbox{ }\Phi_2 = b^{k_1 - k_3} \hbox{ }\Phi_2 \hbox{ }\Phi_{k_1,k_2,k_3} = \Phi_2 \hbox{ }\Phi_{k_1,k_2,k_3} \, ,\nonumber\\
\Phi_{k_1,k_2,k_3}\hbox{ }\Phi_3 = b^{k_2 - k_1} \hbox{ }\Phi_3 \hbox{ }\Phi_{k_1,k_2,k_3} = \Phi_3 \hbox{ }\Phi_{k_1,k_2,k_3} \, .
\end{eqnarray}  
Since $b^n=1$, the center of the algebra is given by the set of $\Phi_{k_1,k_2,k_3}$ such that $k_1=k_2=k_3 \, 
{\rm mod} \, n$.

The generators of the center of the algebra are: $\Phi_{n,0,0},\Phi_{0,n,0},\Phi_{0,0,n},\Phi_{1,1,1}$. 
We call them $x,y,w,z$ respectively. They satisfy the equation
\begin{equation}\label{C3ZnZneq}
x y w = z^n
\end{equation}
which defines the variety $\mathbb{C}^3/\mathbb{Z}_n \times \mathbb{Z}_n$. 
To see this, take $\mathbb{C}^3$ with coordinate $Z^1,Z^2,Z^3$, and consider the action of the group $\mathbb{Z}_n \times \mathbb{Z}_n$ on  $\mathbb{C}^3$ 
\begin{equation}
Z^1,Z^2,Z^3 \rightarrow Z^1 \delta^{-1}\, ,\, Z^2 \delta \xi \,,\, Z^3 \xi^{-1}
\end{equation}
with $\delta^n=\xi^n=1$. The basic invariant monomials under this action 
are $x=(Z^1)^n, y=(Z^2)^n, w=(Z^3)^n, z=Z^1 Z^2 Z^3$ and they clearly satisfy 
the equation (\ref{C3ZnZneq}). 

\begin{figure}[ht]
\begin{center}
\includegraphics[scale=0.45]{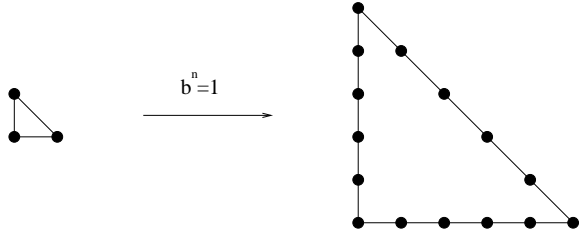} 
\caption{$\mathbb{C}^3 \rightarrow \mathbb{C}^3/\mathbb{Z}_n \times \mathbb{Z}_n$ in the toric picture, $b^5=1$.}
\label{C3ZnZn}
\end{center}
\end{figure}

This fact can be represented in a diagrammatic way as in Figure \ref{C3ZnZn}. This representation of the 
rational value $\beta$-deformation is valid for every toric CY singularity. 

\subsection{Conifold}
The case of the conifold is a bit more intricate and can be a useful example for the generic CY toric cone. The generators of the mesonic algebra 
$x, y, z, w$ satisfy the equations (\ref{comm}). It follows that we can write 
the generic monomial element of the algebra in the ordered form
\begin{equation}\label{mescon}
\Phi_{k_1,k_2,k_3,k_4}=x^{k_1}y^{k_2}w^{k_3}z^{k_4} \, .
\end{equation}
The centre of the algebra is given by the subset of the operators (\ref{mescon}) that satisfy the equations
\begin{eqnarray}
\Phi_{k_1,k_2,k_3,k_4}\hbox{ }x = &b^{k_4 - k_3} \hbox{ }x \hbox{ }\Phi_{k_1,k_2,k_3,k_4}& = x \hbox{ }\Phi_{k_1,k_2,k_3,k_4}\, , \nonumber\\
\Phi_{k_1,k_2,k_3,k_4}\hbox{ }y = &b^{k_3 - k_4} \hbox{ }y \hbox{ }\Phi_{k_1,k_2,k_3,k_4}& = y \hbox{ }\Phi_{k_1,k_2,k_3,k_4} \, ,\nonumber\\
\Phi_{k_1,k_2,k_3,k_4}\hbox{ }w = &b^{k_1 - k_2} \hbox{ }w \hbox{ }\Phi_{k_1,k_2,k_3,k_4}& = w \hbox{ }\Phi_{k_1,k_2,k_3,k_4} \, ,\nonumber\\ 
\Phi_{k_1,k_2,k_3,k_4}\hbox{ }z = &b^{k_2 - k_1} \hbox{ }z \hbox{ }\Phi_{k_1,k_2,k_3,k_4}& = z \hbox{ }\Phi_{k_1,k_2,k_3,k_4} \, . 
\end{eqnarray}  
Because $b^n=1$, the elements of the centre of the algebra are the subset of the operators of the form (\ref{mescon}) 
such that $k_1=k_2$, $k_3=k_4$, mod $n$. 

The centre is generated by 
$\Phi_{n,0,0,0},\Phi_{0,n,0,0},\Phi_{0,0,n,0},\Phi_{0,0,0,n},
\Phi_{1,1,0,0},\Phi_{0,0,1,1}$; we call them respectively $A,B,C,D,E,G$.
The F-term relation
\begin{equation}\label{Fconb}
x y = b w z 
\end{equation}
then implies that $E$ and $G$ are not independent: $E= b G$. 
Moreover the generators of the centre of the algebra satisfy the equations
\begin{equation}\label{orbcon}
A B = C D = E^n \, .
\end{equation}

As in the previous example, it is easy to see that these are the equations of the 
$\mathbb{Z}_n \times \mathbb{Z}_n$ orbifold of the conifold.
Take indeed the coordinates $x,y,w,z$ defining the conifold as a quadric 
embedded in  $\mathbb{C}^4$. 
The action of $\mathbb{Z}_n \times \mathbb{Z}_n$ is
\begin{equation}
x, y, w, z \rightarrow x \delta \,  ,\,  y \delta^{-1}\,  ,\, w \xi^{-1}, z \xi \, ,
\end{equation}
where $\delta ^n = \xi ^n =1$.
The basic invariants of this action are $A,B,C,D,E,G$, and they are subject to the constraint (\ref{Fconb}). Hence the equations (\ref{orbcon}) define the variety $C(T^{1,1})/\mathbb{Z}_n \times \mathbb{Z}_n$. 

\begin{figure}[ht]
\begin{center}
\includegraphics[scale=0.45]{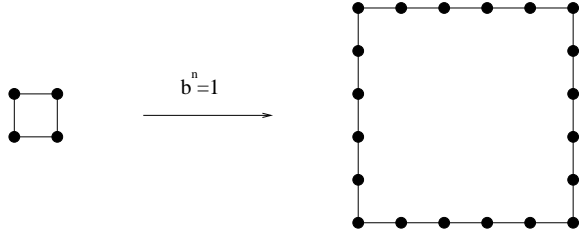} 
\caption{$C(T^{1,1})\rightarrow C(T^{1,1})/\mathbb{Z}_n \times \mathbb{Z}_n$ in the toric picture, $b^5=1$}
\label{T11ZnZn}
\end{center}
\end{figure}

\subsection{The general case}

Now we want to analyse the generic case and show that
 the centre of the mesonic algebra 
for the rational $\beta$-deformed ($b^n=1$) gauge theory is the $\mathbb{Z}_n \times \mathbb{Z}_n$ 
quotient of the undeformed CY.

For a generic toric quiver gauge theory we take a set of basic mesons
$M_{W_j}$ (we will call them simply $x_j$ from now on) corresponding
to the generators $W_j$ of the cone $\mathcal{C}^*$. These are the
generators of the mesonic chiral ring of the given gauge theory.
Because they satisfy the relations (\ref{mb}) it is always possible to
write the generic monomial element of the mesonic algebra generated by
$x_j$ in the ordered form
\begin{equation}\label{genb}
\Phi_{p_1,...,p_k}= x_1^{p_1} x_2^{p_2}...x_k^{p_k} \, .
\end{equation}
We are interested in the operators that form the centre of the
algebra, or, in other words, that commute with all the elements of the
algebra.  To find them it is enough to find all the operators that
commute with all the generators of the algebra, namely
$x_1,...,x_k$. The generic operator (\ref{genb}) has charge
$Q_{p_1,...,p_k}$ under the two flavour $U(1)$ symmetries, and the
generators $x_j$ have charges $Q_j$. They satisfy the following
relations
\begin{equation}\label{relc}
\Phi_{p_1,...,p_k} \hbox{ }x_j= x_j \hbox{ }\Phi_{p_1,...,p_k} \hbox{ }b^{Q_{p_1,...,p_k} \wedge Q_j } \, .
\end{equation}
This implies that the centre of the algebra is formed by the set of $\Phi_{p_1,...,p_k}$ such that
\begin{equation}
Q_{p_1,...,p_k} \wedge Q_j = 0 \hbox{ mod $n$} \hbox{ , } {  j=1,...,k} \, .
\end{equation}
At this point it is important to realize that the $Q_j$ contain the
two dimensional vectors perpendicular to the edges of the two
dimensional toric diagram.  The fact that the toric diagram is convex
implies that the $Q_j$ span the $T^2$ flavour torus.  In particular
the operator $\Phi_{p_1,...,p_k}$ must commute (modulo $n$) with the
operators with charges $(1,0)$ and $(0,1)$. The first condition gives
all the operators in the algebra that are invariant under the
$\mathbb{Z}_n$ in the second $U(1)$, while the second gives all the
operators invariant under the $\mathbb{Z}_n$ contained in the first
$U(1)$.  All together the set of operators in the centre of the
algebra consists of all operators $\Phi_{p_1,...,p_k}$ invariant under
the $\mathbb{Z}_n \times \mathbb{Z}_n$ discrete subgroup of the $T^2$.

The monomials made with the free $x_1,...,x_k$ coordinates of
$\mathbb{C}^k$ that are invariant under $\mathbb{Z}_n \times
\mathbb{Z}_n$, form, by definition, the quotient variety
$\mathbb{C}^k/\mathbb{Z}_n \times \mathbb{Z}_n$. The toric variety
$V$ is defined starting from a ring over $\mathbb{C}^k$ with relations
given by a set of polynomials $\{ q_1,...,q_l \}$ defined by the toric
diagram
\begin{equation}
\mathbb{C}[V]= \frac{\mathbb{C}[x_1,...,x_k]}{\{q_1,...,q_l \}} \, .
\end{equation} 
Indeed the elements of the centre of the algebra are the monomials made with 
the $x_j$, subject to the relations $\{q_1,...,q_l \}$, invariant under $\mathbb{Z}_n \times \mathbb{Z}_n$. 
This fact allows us to conclude that the centre of the algebra in the case $b^n=1$ is the quotient of the original $CY$
\begin{equation}
V_b= \frac{{\rm CY}}{\mathbb{Z}_n \times \mathbb{Z}_n} \, .
\end{equation} 
The $\beta$-deformed $\mathcal{N}=4$ gauge theory and the $\beta$-deformed conifold gauge theory 
are special cases of this result. In the appendix we will discuss 
a more sophisticated example, which includes $SPP$ as a particular case.

\section{Conclusions}
In this paper we discussed general properties of the $\beta$-deformation of
toric quiver gauge theories and of their gravitational duals, which have
a very simple characterization in terms of generalised complex geometry.

We analysed the moduli space of vacua of the $\beta$-deformed 
theory using D-branes probes and field theory analysis. An important class of 
supersymmetric probes, the giant gravitons, has still to be analysed. It would
be interesting to study the classical configurations of giant gravitons
in the $\beta$-deformed background and their quantisation. This should
give information about the spectrum of BPS operators and, as it happens
in the undeformed theory, it should help in computing partition functions
for the chiral ring of the gauge theory \cite{Minwalla,FHZ,Forcella:2007wk,BVFHZ,Forcella:2007ps,benve,Hanany:2006uc,Feng:2007ur}.

On the gravity side, we clarified the geometrical structure of the 
supersymmetric $\beta$-deformed background. The description in terms of pure
spinors is remarkably simple. It would be interesting to see whether 
this description can be extended to the analysis of other marginal deformations
of superconformal theories. In particular ${\cal N}=4$ SYM and other quiver
gauge theories admit deformations that breaks the $U(1)^3$ symmetry whose
supergravity dual is still elusive. It would be interesting to extend
our methods to the search of these missing solutions.

\section*{Acknowledgments}

D.~F.~ would like to thank Loriano Bonora, Umut Gursoy, Alberto
Mariotti, Marco Pirrone for valuable discussions.  D.~F.~ would like
to thank the Laboratoire de Physique Th\'eorique de l'\'Ecole Normale
Sup\'erieure and Universit\'es Paris VI et VII, Jussieu for the kind
hospitality during part of this work. A.Z. would like to thank the
Laboratoire de Physique Th\'eorique de l'Universit\'es Paris VI et
VII, Jussieu, the Galileo Galilei Institute in Florence and the Newton
Institute in Cambridge for hospitality and support during part of this
work.  D.~F.~ is supported in part by INFN and the Marie Curie
fellowship under the programme EUROTHEPHY-2007-1. A.~Z.~ is supported
in part by INFN and MIUR under contract 2005-024045-004 and
2005-023102 and by the European Community's Human Potential Program
MRTN-CT-2004-005104. A.B and M.P. are supported in part by the RTN contract 
MRTN-CT-2004-512194 and by ANR grant BLAN05-0079-01. R.M. 
is supported in part by RTN contract  MRTN-CT-2004-005104 and  
by ANR grant BLAN06-3-137168. L. M. is supported by the DFG cluster
of excellence ``Origin and Structure of the Universe" and 
would like  to acknowledge the Galileo Galilei Institute for
Theoretical Physics for hospitality and the INFN for partial support.

\begin{appendix}

\section{$\beta$-deformed ${\cal N}=4$ Super Yang-Mills}\label{apxa}
For the $\beta$-deformation of ${\cal N}=4$ SYM it is possible to use the
pure spinor formalism to determine the precise relation between the
parameter $\gamma$ entering the supergravity 
background and the $\beta$ parameter deforming the superpotential of
the dual gauge theory. Even if the computation does not apply to the
$\beta$-deformation of a generic toric Calabi-Yau, we report it here
since it provides a nice application of the formalism of Generalised
Complex Geometry.

The computation is based on the observation that for a generic
deformation of ${\cal N}=4$ SYM it possible to relate the integrable pure
spinor of the gravity solution ($\hat{\Psi}_-$ for us) and the
superpotential of the dual gauge theory \cite{lucasup, mpz}.  More
precisely it possible to write the superpotential for a single D-brane
probe, with a world-volume flux ${\rm F}$ and wrapping a cycle
$\Sigma$ in the internal manifold, in terms of the closed pure spinor
\cite{lucasup}. Since $ e^{ 3 A}\hat\Psi_-$ is closed, one can locally write
$ e^{3 A}\hat\Psi_-=\d\chi$ and the superpotential can be written as
\bea\label{branesup}
\calw=\int_\Sigma \chi|_\Sigma\wedge e^{\text{F}}\ .
\eea


Notice that (\ref{branesup}) has precisely the form of the CS term in the
standard D-brane action, where $\chi$ plays the role of the twisted
RR-potentials $C\wedge e^B$. A non-abelian generalisation of such CS
term for multiple D-branes was obtained by Myers in \cite{myers},
using an argument essentially based on T-duality. Since the pure
spinor $\hat\Psi_-$ transforms precisely as the RR-field strengths
under T-duality, the same argument can be applied in our case, and the
resulting non-abelian superpotential has exactly the same form of
Myers' non-abelian CS term, with $C\wedge e^B$ substituted by $\chi$.

For the background obtained by $\beta$-deforming $AdS_5\times S^5$, using the standard 
flat complex coordinates on the internal warped $\mathbb{C}^3$, we have
\bea
e^{3 A}\hat\Psi_-=\gamma(z^1z^2\d z^3 + \text{ cyclic} )+\d z^1\wedge\d z^2\wedge \d z^3\ ,
\eea 
and thus
\bea
\chi=\gamma z^1z^2z^3+\frac{1}{3!}\epsilon_{ijk}z^i\d z^j\wedge \d z^k\ .
\eea  
Then, from the above argument and Myers' non-abelian CS action we get
the following non-abelian superpotential for a stack of D3-branes (in units $\alpha^\prime=1$)
\bea
\calw&=&{\rm Str}[e^{2 i\pi \iota_\Phi\iota_\Phi}\chi]_{(0)}\cr
&\sim&\tr[(1+i \pi \gamma)\Phi_1 \Phi_2 \Phi_3-(1- i \pi \gamma)\Phi_1 \Phi_3 \Phi_2]\ ,
\eea
where $\Phi_i$ is the non-abelian scalar field describing the D3-brane
fluctuations, which is canonically associated to $z^i/(2 \pi \alpha')$. Comparing
with (\ref{N=4beta}), since we need $\gamma\ll 1$ to trust the
supergravity approximation, we conclude that
\bea
\beta=\gamma\ .
\eea

\section{Some explicit field theory examples}\label{apx}
In this appendix we illustrate few
points of the field theory analysis. Using the $SPP$ example, we show how the
non commutative product acts on the undeformed superpotential and motivate
formula (\ref{defSup}). We also discuss the non abelian branches of the 
theories $L^{p,q,q}$ for rational $\beta$.    

\subsection{Action of the non commutative product}
To obtain the $\beta$-deformed gauge theory we pass from the simple product between fields to the star product:
\begin{equation}\label{starF}
X_i X_j \rightarrow X_i * X_j \equiv e^{i \pi \beta (Q^i \wedge Q^j)} X_i X_j 
\end{equation}  
where $X_i$ are the elementary fields in the quiver.

The star product is non commutative but associative and the product of a string of $n$ fields takes the form:
\begin{equation}\label{starFs}
X_{a_1}*...*X_{a_n} \equiv b^{-1/2 (\sum_{i<j}Q_{a_i} \wedge Q_{a_j})} X_{a_1}...X_{a_n}
\end{equation} 
Let us consider two generic mesonic fields with base point in the same
gauge group: $M=X_{a_1} \ldots X_{a_m}$, $N=X_{b_1} \ldots
X_{b_n}$. In the undeformed theory they commute $M N=N M$, but when we
turn on the $\beta$-deformation this relation becomes: $\tilde{M} *
\tilde{N} = \tilde{N} * \tilde{M}$, for the quantities
$\tilde{M}=X_{a_1} * \ldots * X_{a_m}$, $\tilde{N}=X_{b_1} * \ldots *
X_{b_n}$. This gives, using (\ref{starFs}):
\begin{equation}
\tilde{M} \tilde{N} = b^{(Q_M \wedge Q_N)} \tilde{N} \tilde{M}
\label{commutationnew}
\end{equation}
where we defined the charges of the composite fields:
$Q_M=Q_{a_1}+...+Q_{a_m}$, $Q_N=Q_{b_1}+...+Q_{b_n}$. Note that
relation (\ref{commutationnew}) also holds in the same form for mesons
$M$ and $N$, since they are proportional to $\tilde M$ and $\tilde N$
respectively, thanks again to (\ref{starFs}). We obtain therefore our
general method (\ref{mesb}) for computing commutation relations for
mesons.

We would like now to understand the structure of the superpotential
$W$ for the $\beta$-deformed theory, obtained by replacing the
standard product with the star product in (\ref{starF}). First of all,
since $W$ is a trace of mesons, consistency requires the star product
to be invariant under cyclic permutations of the fields. This happens
because of the conservation of charge \footnote{This is the analog of
  the cyclic invariance of the factor $\exp \Big(-\frac{i}{2}
  \theta_{ij} \sum_{0<\mu < \nu <n} k_{\mu}^i\hbox{ } k_{\nu}^j \Big)$
  in the $n$ point vertex interaction of the perturbative expansion of
  space-time non-commutative quantum field theories, due to the
  conservation of momenta at each vertex.}: the two $U(1)$ flavour
charges of each meson are zero.

Then we want to show that $W$ can always be put into the form
(\ref{defSup}) by rescaling fields.
Consider a generic toric gauge theory with $G$ gauge groups, $E$ elementary
fields and $V$ monomials in the superpotential. We have the relation
\cite{dimers}:
\begin{equation}
G-E+V=0
\label{euler}
\end{equation}

The superpotential $W$ of the undeformed theory is a sum of $V$ monomials
$m_I,n_J$ made with traces of products of elementary fields. Every
elementary field appears in the superpotential $W$ once with the
positive sign and once with the negative sign,
\begin{equation}\label{superpotnb}
W=\sum_{I=1}^{V/2} c_I^+ m_I - \sum_{J=1}^{V/2} c_J^- n_J
\end{equation}
After $\beta$-deformation the coefficients $c_I^+$, $c_J^-$ are
replaced by generic complex numbers.

Rescaling the elementary chiral fields produces a rescaling also of
the coefficients $c_I^+$, $c_J^-$, but note that the quantity
\begin{equation}\label{cost}
\frac{\prod_I c_I^+ }{\prod_J c_J^-}= \hbox{ const }
\end{equation}
remains constant since every chiral field contributes just once in the
numerator and just once in the denominator. In the undeformed theory
this constant is 1, while in the $\beta$-deformed case its value can
be written as $b^{-\alpha V/2}$, for some rational $\alpha$.

Consider the action of the $E$ dimensional group of chiral fields
rescalings over the $V$ dimensional space of coefficients $c_I^+$,
$c_J^-$ in the superpotential. The subgroup that leaves invariant a
generic point (with all coefficients different from zero) is the group
of global symmetries of the superpotential. It is known that toric
theories have $G+1$ global symmetries\footnote{These are the $2$
  flavour non anomalous symmetries plus $G-1$ baryonic symmetries
  (anomalous and non anomalous).}, therefore the dimension of a
generic orbit is $E-(G+1)=V-1$, thanks to (\ref{euler}). This shows
that (\ref{cost}) is the only algebraic constraint under field
rescalings, and hence it is always possible to put the superpotential
in the form:


\begin{equation}\label{superpotb2}
W=\sum_{I}m_I - b^{\alpha} \sum_{J} n_J
\end{equation}

Let us explain in more detail a particular case, $SPP$.

\begin{figure}[ht]
\begin{center}
\includegraphics[scale=0.50]{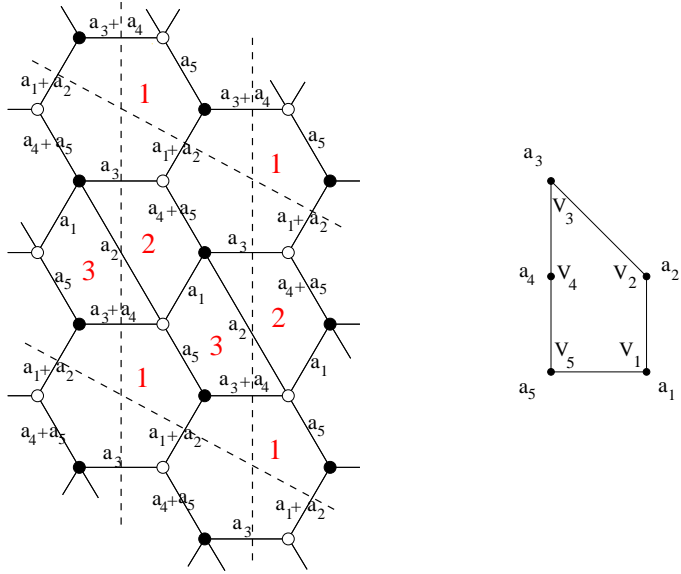} 
\caption{Dimer configuration and toric diagram for the $SPP$ singularity.}
\label{sppf}
\end{center}
\end{figure}
 

All the information of a toric quiver gauge theory is encoded in a
dimer graph \cite{dimers} (see Figure \ref{sppf}).  The idea is very
simple: you draw a graph on $T^2$ such that it contains all the
information of the gauge theory: every link is a field, every node a
superpotential term, and every face is a gauge group. There exist
efficient algorithms to compute the distribution of charges $a_i$ for
the various $U(1)$ global symmetries of the gauge theory
\cite{aZ}. The charges for every fields in the $SPP$ gauge theory are
given in Figure \ref{sppf}. For the two global flavour symmetries we
are interested in, the trial charges are such that $\sum_i a_i=0$
(conservation of flavour charges at every node).  We can thus write
the charges of the mesonic fields in terms of the trial charges:
\begin{eqnarray}
& & x = X_{11} \rightarrow a_1+a_2 \hbox{ , } y= X_{12} X_{21} \rightarrow a_3+ a_4 + a_5\nonumber\\
& & w = X_{13} X_{32} X_{21} \rightarrow a_2+ 2 a_3 +a_4 \hbox{ , } z=  X_{12} X_{23} X_{31} \rightarrow a_1 + a_4 + 2 a_5 \nonumber\\
\end{eqnarray} 
Using the values of the mesonic charges given in (\ref{messppc}) one
can now compute the charges $a_i$ for the elementary fields. These
will be a set of rational numbers. We can now use these charges to
pass from the simple product to the star product (\ref{starF}) in
every term in the superpotential. This procedure will generate a phase
factor in front of every term in the superpotential. The interesting
quantity is the invariant constant in (\ref{cost}):
\begin{equation}\label{costspp}
\frac{\prod_I c_I^+ }{\prod_J c_J^-}= e^{2 i \pi \beta}=b^{-1}
\end{equation}
The actual value of this constant implies that we can rescale the
elementary fields in such a way that the superpotential assumes the
form:
\begin{equation}\label{superpotb}
W=X_{21}X_{12}X_{23}X_{32} + X_{13}X_{31}X_{11} - b^{1/2}( X_{32}X_{23}X_{31}X_{13} + X_{12}X_{21}X_{11} )
\end{equation}
Using the F-term equations from the $\beta$-deformed superpotential
(\ref{superpotb}) one can reproduce the commutation rules among mesons
(\ref{sppbc}) given in the main text plus the $\beta$-deformed version
of the CY singularity: $w z = b x y^2$.

\subsection{$L^{p,q,q}$}
\begin{figure}[ht]
\begin{center}
\includegraphics[scale=0.55]{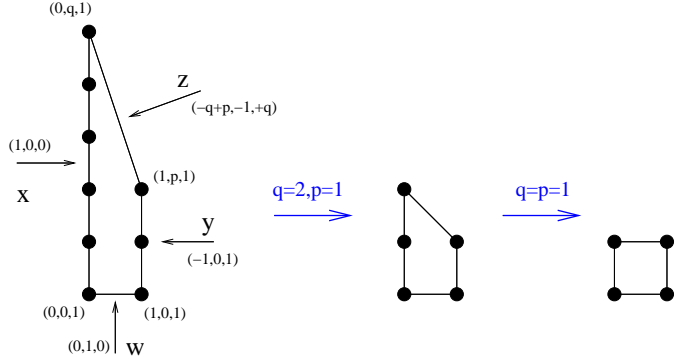} 
\caption{The toric diagrams of the $C(L^{p,q,q})$ singularity and their two well known special cases: $SPP$, $C(T^{1,1})$.}
\label{Lpqq}
\end{center}
\end{figure}

In this Section we give another example of the moduli space for
rational $\beta$.  $L^{p,q,q}$ with $q \ge p$ are an infinite class of
Sasaki-Einstein spaces. For some values of $p,q$ these spaces are very
well known. Indeed $L^{1,1,1}=C(T^{1,1})$, and $L^{1,2,2}=\hbox{$SPP$}$.
The real cone over $L^{p,q,q}$ is a toric Calabi-Yau cone that can be
globally described as an equation in $\mathbb{C}^4$:
\begin{equation}\label{lpqqeq}
C(L^{p,q,q}) \rightarrow x^p y^q= w z
\end{equation}
All the algebraic geometric information regarding these singularities
can be encoded in a toric diagram, see Figure \ref{Lpqq}.
The variety is a complete intersection in $\mathbb{C}^4$. Indeed to
each generator of the dual cone we can assign a coordinate like in
Figure \ref{Lpqq}.  These coordinates are in one to one correspondence
with the mesonic field in the field theory generating the chiral ring,
and the first two coordinates of the vectors are their charges under
the two $U(1)$ flavour symmetries.  The generators of the mesonic
algebra are $x,y,w,z$ and thanks to their commutation relations
\begin{eqnarray}
& & x y = y x \hbox{ , } x w = b w x \hbox{ , } x z = b^{-1} z x \nonumber\\
& & y w = b^{-1} w y \hbox{ , } y z = b z y \hbox{ , } w z = b^{q-p} z w
\end{eqnarray}
we can write the generic monomial element of the algebra in the ordered form:
\begin{equation}\label{meslpqq}
\Phi_{k_1,k_2,k_3,k_4}=x^{k_1}y^{k_2}w^{k_3}z^{k_4} 
\end{equation}
The center of the algebra is given by the subset of the operators (\ref{meslpqq}) that satisfy the equations:
\begin{eqnarray}
\Phi_{k_1,k_2,k_3,k_4}\hbox{ }x = &b^{k_4 - k_3} \hbox{ }x \hbox{ }\Phi_{k_1,k_2,k_3,k_4}& = x\hbox{ } \Phi_{k_1,k_2,k_3,k_4} \nonumber\\
\Phi_{k_1,k_2,k_3,k_4}\hbox{ }y = &b^{k_3 - k_4} \hbox{ }y \hbox{ }\Phi_{k_1,k_2,k_3,k_4}& = y\hbox{ } \Phi_{k_1,k_2,k_3,k_4} \nonumber\\
\Phi_{k_1,k_2,k_3,k_4}\hbox{ }w = &b^{k_1 - k_2 -(q-p) k_4}\hbox{ } w \hbox{ } \Phi_{k_1,k_2,k_3,k_4}& = w \hbox{ } \Phi_{k_1,k_2,k_3,k_4} \nonumber\\ 
\Phi_{k_1,k_2,k_3,k_4} \hbox{ }z = &b^{k_2 - k_1+(q-p)k_3}\hbox{ } z \hbox{ }\Phi_{k_1,k_2,k_3,k_4}& = z \hbox{ }\Phi_{k_1,k_2,k_3,k_4} \nonumber\\ 
\end{eqnarray}  
Because $b^n=1$ the elements of the center of the algebra are the subset of the operators of the form (\ref{meslpqq}) 
such that $k_3=k_4$, $k_1 = k_2 +(q-p) k_4 $, $k_1= k_2 + (q-p) k_3 $ mod $n$. \\
The generators of this algebra are $\Phi_{n,0,0,0},\Phi_{0,n,0,0},\Phi_{0,0,n,0},\Phi_{0,0,0,n},
\Phi_{1,1,0,0},\Phi_{q-p,0,1,1}$; we call them respectively $A,B,C,D,E,G$.
Using the F-term relation $x^p y^q = w z$ we see that $G$ depends on the other generators through: $G=E^q$. Moreover the relations among generators are: 
\begin{equation}\label{quotLpqq}
A^p B^q = C D \, , \qquad E^n= A B \, .
\end{equation}
In the special case of $q=p=1$ these equations reduce to those for the
quotient of the conifold.  It is easy to see that equations
(\ref{quotLpqq}) define the $\mathbb{Z}_n \times \mathbb{Z}_n$
orbifold of the $C(L^{p,q,q})$. Take the coordinates $x,y,w,z$
realizing $C(L^{p,q,q})$ as a quadric embedded in $\mathbb{C}^4$. The
action of $\mathbb{Z}_n \times \mathbb{Z}_n$ is:
\begin{equation}\label{eqquoL}
x, y, w, z \rightarrow x \delta , y \delta^{-1} , w \xi , z \delta^{-q+p}\xi^{-1}
\end{equation}
where $\delta ^n = \xi ^n =1$.  The independent invariants of this
action are $A,B,C,D,E$, and they are subject to the constraints
(\ref{quotLpqq}). Hence the equations (\ref{quotLpqq}) define the
variety $C(L^{p,q,q})/\mathbb{Z}_n \times \mathbb{Z}_n$.

\end{appendix}

\end{document}